\documentclass[a4paper,twoside]{article}

\newcommand{\affil}[1]{$^{\rm #1}$}
\usepackage[authoryear]{natbib}
\bibpunct{(}{)}{;}{a}{}{,}
\usepackage{graphicx}
\date{} 
\title{\large\bf\flushleft KOI-256's Magnetic Activity under the Influence of the White Dwarf}
\author{\parbox{\textwidth}{\flushleft
\vspace{-0.5cm}
{\it Ezgi YOLDA\c{S}\affil{1},\affil{2}, Hasan Ali DAL\affil{1}}\\
\vspace{0.4cm}
{\small \affil{1}\,Department of Astronomy and Space Sciences, University of Ege, Bornova, 35100 ~\.{I}zmir, Turkey}\\
{\small \affil{2}\,Corresponding Author, Email: ezgiyoldas@gmail.com}}}
\begin{document}
\twocolumn[
\begin{changemargin}{.8cm}{.5cm}
\begin{minipage}{.9\textwidth}
\vspace{-1cm}
\maketitle
%
%
\small{\bf Abstract:} We present the findings about chromospheric activity nature of KOI-256 obtained from the Kepler Mission data. Firstly, it was found that there are some sinusoidal variations out-of-eclipses due to cool spot activity. The sinusoidal variations modelled by the SPOTMODEL program indicate that the active component has two different active regions. Their longitudinal variation revealed that one of them has a migration period of 3.95 years, while the other has a migration period of 8.37 years. Secondly, 225 flares were detected from the short cadence data in total. The parameters, such as increase ($T_{r}$) and decay ($T_{d}$) times, total flare time ($T_{t}$), equivalent durations ($P$), were calculated for each flare. The distribution of equivalent durations versus total flare times in logarithmic scale is modelled to find flare activity level. The $Plateau$ value known as the saturation level of the active component was calculated to be 2.3121$\pm$0.0964 s, and the $Half-life$ value, which is required flare total time to reach the saturation, was computed to be 2233.6 s. In addition, the frequency of $N_{1}$, which is the number of flares per an hour in the system, was found to be 0.05087 $h^{-1}$, while the flare frequency $N_{2}$ that the flare-equivalent duration emitting per an hour was found to be 0.00051. Contrary to the spot activity, it has been found that the flares are in tends to appear at specific phases due to the white dwarf component.\\

\medskip{\bf Keywords:}techniques: photometric - methods: data analysis - methods: statistical - binaries: eclipsing - stars: low-mass - stars: flare - stars: individual (KOI-256).

\medskip
\medskip
\end{minipage}
\end{changemargin}
]
\small

\section{Introduction}

65$\%$ of the stars in our galaxy are red dwarf stars, and 75$\%$ of them show flare activity \citep{Rod86}. The vast majority of the red dwarfs found in the open star clusters and association show flare activity \citep{Mir90, Pig90}, while the number of UV Ceti stars in the clusters decrease from a young cluster to the older one \citep{Sku72, Mar92, Pet91, Sta91}. The flare activity results in mass loss, which plays an important role in stellar evolution.

Although various studies have been carried out since the first flare was observed by R. C. Carrington and R. Hodgson on September 1, 1859 \citep{Car59, Hod59}, it is not fully understood what the flare and its process are. However, the flare activity of the dMe stars is modelled on the basis of the Solar Flare Event \citep{Ger05, Ben10}. The studies, which have been continuing to understand the flare events of dMe stars have revealed that there are some differences between the flare energy levels of stars \citep{Ger83}.

The level of highest energy releasing on the Sun are about $10^{30}$ - $10^{31}$ erg in the flare events \citep{Ger05, Ben08}. It seems that this level is about $10^{31}$ erg \citep{Hai91}, if RS CVn stars, the chromospheric active binaries, are considered. The observations lasting over decades show that the energy levels of flares occurring on dMe stars could have increased from $10^{28}$ erg to $10^{34}$ erg \citep{Ger05}. Considering the stars in the Pleiades cluster and Orion association, it is seen that these values have reached $10^{36}$ erg \citep{Ger83}. There are significant differences between the energy level of flare stars from different spectral types. For example, it is well known that there are serious differences between the mass loss rates and the flare energy level, if the Sun is compared to a dMe star. Nevertheless, the flare events occurring on the dMe stars are tried to be explained with the Solar Flare processes. As a result, it is clear that the flares in the different stars should be well studied and the similarities and differences between them should be analyzed. \citet{Ger05} and \citet{Hud97} have suggested that magnetic reconnection processes are the source of energy for flare events. In order to determine the important points in the flare process, it is necessary that the cause of the differences in the flare frequency and also the flare energy spectra should be determined.

Besides \citet{Ger83}, there are several studies about the flare energies in the literature such as \citet{Ger72, Lac76, Wal81, Pet84, Mav86}. However, according to \citet{Dal10, Dal11}, the results obtained in these studies are unsatisfactory for comparing the stars from different spectral types. For example \citet{Ger72} derived the flare energy spectra for several dMe stars, such as AD Leo, EV Lac, UV Cet and YZ CMi, while \citet{Ger83} derived the flare energy spectra for lots of stars from the galactic field to compare with the flare energy spectra of some stars from Pleiades and Orion association. They showed that these stars are located in different points in the distribution of energy. They correctly indicated that this distribution is caused due to different ages of the stars. On the other hand, \citet{Dal10, Dal11} indicated that there is one more reason to cause this distribution. The second reason is including the luminosity parameter in calculations, of flare energies, which leads to incorrect results, as this caused the stratifying in the flare energy spectra. Indeed, the studies of \citet{Dal10, Dal11} show that the flare equivalent durations in the logarithmic scale vary within a certain rule versus the flare total time, and it depends on the spectral type of the stars, on which the flares occur, when the relations between the flare parameters are examined. On the other hand, two different flare frequencies were defined by \citet{Ish91} for the flare activity. The frequency of $N_{1}$ indicates the number of flares per an hour, while the frequency of $N_{2}$ describes the flare-equivalent duration emitting per an hour. \citet{Let97} clearly show that the frequency of EV Lac's flares vary over time.

On the other hand, for the first time in the literature, \citet{Kro52} discovered that UV Ceti-type stars also exhibited spot activity known as BY Dra Syndrome. \citet{Kro52} observed sinusoidal variations out-of-eclipses in the light curves of the YY Gem, which is a binary system, and \citet{Kun75} called this phenomenon BY Dra Syndrome and explained as a fact that these variations were caused by heterogeneous temperature distribution on the surface of the star. In the case of the Sun, \citet{Ber05} found two stable active longitudes separated by 180 degrees from each other, and they indicated that these longitudes are exhibiting semi-rigid behavior. However, according to some authors, these longitudes are not the persistent active structures, which show variation in the time \citep{Lop61, Sta72, Bog82}. The difference between the regular activity oscillations shown by these longitudes, called the Flip-Flop, is very important for the north-south asymmetry exhibited by the star's magnetic topology. It is very important to calculate the angular velocities of these longitudes because these calculations light on the differential rotational velocities in the latitudes of the spots and spots groups.

In the case of binary or single stars, determining the parameters of stellar spot activity, such as spot latitude, radius and longitude, is a controversial phenomenon. In the literature, there are several methods to find out these parameters \citep{Rib02, Rib03, Wal13a, Jef09}. For example, the SPOTMODEL program \citep{Rib02, Rib03} required two band observation and the inclination of the rotation axis to be able to model the distribution of the spots on the stellar surface. However, it must be noted that the system's maximum brightness level ever observed has also a very important role to determine the spot radius depending on spot latitude. According to \citet{Wal13a, Jef09}, this method does not work for the observations such as the data acquired in the Kepler Mission. Because the Kepler observations provide us monochromatic data, which have been detrended while combining different observation parts.

In this study, we examine the flare and spot activity exhibiting by KOI-256, and examine the parameters derived from the One Phase Exponential Association (hereafter OPEA) model, and also the spot migration considering the sinusoidal variation out-of-the flare. It is seen from the literature that KOI-256 is a system, which the primary component is a white dwarf, the secondary component is a main sequence star from the spectral type of M3. In addition, there are some clues about third body, which could be a planet. \citet{Bor11} indicated that KOI-256 is a planet candidate system. There are lots of studies such as \citet{Rit03, Bor11, Sla11, Wal13b, Mui13, Mui14, Zac04} in the literature. Several physical parameters of KOI-256 were computed in these studies, using some colour calibrations explained by these authors. As it can be seen from these parameters listed in Table 1, KOI-256 is very interesting system. Although one of the components is a with dwarf, \citet{Wal13b} indicated that the system's age is 0.01 Gyr. However, using the Equation 2.3 given by \citet{Gan97} with the stellar parameters taken from \citet{Mui13}, we estimated the age of the system as $\sim$2 Gyr, considering the cooling of the white dwarf component. Indeed, \citet{Mui13} laterly revealed that the system is a post-common envelope binary. In the paper, the stellar spot activity analyses are described in Section 2.1, while the flare models are described in Section 2.2. The orbital period variation analysis is explained in Section 2.3. The results obtained from the analyses are summarised and discussed in Section 3.

\section{Data and Analyses}

The Kepler Mission project, which was launched to explore the exoplanets, has observed more than 150.000 sources \citep{Bor10, Koc10, Cal10}. These observations are one of the highest sensitivity photometric observations ever achieved \citep{Jen10a, Jen10b}. With this highest sensitivity of observation, a large number of new eclipsing binaries and lots of new variable stars have been discovered besides the exoplanets \citep{Sla11, Mat12}. Some of these variable stars are single, and some of them are binary stars, which exhibit both the stellar spot activity and flare activity \citep{Bal15}. In this study, observational data of KOI-256, which is one of these binary systems, was taken from Kepler Data Base \citep{Sla11, Mat12}. KOI-256's short cadence data obtained in the Kepler Mission covered the time range from BJD 24 55372.460219 to BJD 24 55552.55836 and from BJD 24 56419.80351 to BJD 24 56424.01160, while the long cadence data of the system covered time ranges BJD 24 54964.51238-55206.21898, BJD 24 55276.51124-55552.54849, BJD 24 55641.51645-55931.30552, BJD 24 56015.77828-56304.13695, BJD 24 56392.24699-56424.00173. All the available data presented in both Long and Short Cadence formats are shown in Figure 1. The short cadence data were used in the analyses of the flares, while the long cadence data were used for the analyses of sinusoidal variation. The detrended data was used among the public data provided in the Kepler Data Base. The data sets were created in appropriate formats, which were edited in the analysis processes for the flare activity, the stellar spot activity and the orbital period variation ($O-C$).

The whole of KOI-256's analysed data taken from the Kepler Database is shown in the upper and middle panels of Figure 1. The light curve is plotted versus the Barycentric Julian Date in the upper panel, while it is plotted versus phase in the middle panel of the figure. In the lower panel, the light curve is plotted versus phase, expanding the y-axis to be easily seen the sinusoidal variations out-of-the dominant flare activity.

\subsection{Stellar Spot Activity}

Examining the out-of-eclipses light variations, it is seen that the system exhibits also sinusoidal variations. Considering the surface temperatures of the components of the system, it is understood that the variations is caused by the rotational modulation of the cool stellar spots. It is seen that both minima times and amplitudes of the light curves are varying once in a few cycles, when the consecutive cycles depend on the orbital period are examined in the light curves, in which both the flares and eclipses are removed. This situation indicates that the active regions on the components of the system evolved rapidly. Because of this, it is not possible to model the entire light curve in a single analysis, so the data of sinusoidal variation are separated into several sets. When dividing into data sets, consecutive cycles in which some characters of the light curve such as a light curve asymmetry, a spot minima phase, and the minima and maxima level were the same were collected in a single set. In this format, 138 sub-sets are obtained, thus, each sub-set is modelled separately.

To be able to model the sinusoidal variations, the pre-whitened light curves were obtained. In this step, we firstly removed data parts, in which all instant light increase due to the flares are seen, from whole data. In addition, the data parts, in which the primary minima are seen due to eclipses, were also removed from whole data.

The pre-whitened light curves are modelled with the SPOTMODEL program \citep{Rib02, Rib03} to find out the spot distribution parameters, such as the spot radius, latitude and especially longitudinal distribution, on the stars. To model the spots, the SPOTMODEL program requires two band observations or a temperature factor ($kw=[T_{spot}/T_{stellar surface}]^2$) for the stellar spot. However, the data analysed in this study consist of monochromatic observations that are presented publicly in the Kepler Database. In this point, considering both the study of \citet{Bot78}, which first revealed the spot temperature factor for the stellar activity, and also the light curve analysis obtained from analogue systems \citep{Cla01, Tho08}, it was assumed that the secondary component exhibits magnetic activity. The inclination of rotation axis is taken 89$^\circ$.1 as it was given by \citet{Mui13}. Then, taking the different values from $kw=0.60$ to $kw=0.90$ for the spot temperature factor in agreement with the values found by these studies for analogue systems, the first few sets were tried to be modelled. As a result, it was seen that the best solution can be obtained by taking the spot temperature factor of $kw=0.65$ for both spots. For this reason, the spot temperature factors were taken as $kw=0.65$ for both spots in the all sets. Taking the spot temperature factor as constant value in the models for each set, the longitudinal, latitudinal distributions and radius variations of the spot area on the active component were determined.

In the analysis, the parameters such as the longitudes ($l$), latitudes ($b$), and radii of the spots ($g$) parameters were taken as free parameters. The parameters obtained from the models for each set are listed in Table 2, while six examples for the derived models are shown in Figure 2. In the left panel of the figure, the observations and models are plotted versus Heliocentric Julian Day as a time, while they were plotted versus phase computed using by orbital period. As seen from the figure, the synthetic models absolutely fit the observations. The variations of latitude ($b$) and radius ($g$) for both spots are shown in Figures 3. In addition, the most important parameter of the models, longitude ($l$), is plotted versus time in Figure 4. The longitudinal variations were fitted by a linear function for both spots. Using these linear fits, the migration periods were computed. The migration period was found to be 3.95 year for first spotted area, and 8.37 year for second area.

To test whether the findings about longitudinal variations are close to the real nature of the stellar surface, it was tested by another method. Using the Fourier Transform, the minima times of sinusoidal variations, where the amplitude is larger, were computed. Then, the orbital phases called as $\theta_{min}$ were computed for these sinusoidal minima times. In figure 5, the variation of $\theta_{min}$ are plotted versus time. This variation was fitted by a linear function similar to the longitudinal variations. In the same process, the migration period was computed from the phase shift of the $\theta_{min}$ using this linear function. In a result, the period of the spotted area migration was found to be 9.126 year. As it is expected, this value is in agreement with one of the migration periods found from the longitudinal variations.

\subsection{Flare Parameters and Models}

In order to understand the nature of the flare activity and to find out the flare behaviour of the system, we tried to determine the flares occurring on the active component. For the reason, using the synthetic light curves, all the other variations apart from the flares were removed from the entire light curve. Since the system is an eclipsing binary, all minima light variations between the phases of 0.04 around the minima points, where two components are external tangent to each other, were removed from the entire light curve. However, the separated observations due to the technical reasons were also removed from the data.

In order to specify the flare parameters, such as start and end points of a flare and its energy, it should be defined the quiet level of the light curve. However, it was seen some sinusoidal variations due to the rotational modulation exhibited by one of the components in addition to the flares. For this reason, the synthetic light curves were derived for the light variations out-of-the flares using the Fourier transform given by Equation (1), and all the data were modelled with these synthetic light curves for the all phases.

\begin{center}
\begin{equation}
\label{eq:five}
L(\theta)= A_{0} ~ + ~ \sum_{\mbox{\scriptsize\ i=1}}^N ~ A_{i} ~ cos(i \theta) ~ + ~ \sum_{\mbox{\scriptsize\ i=1}}^N ~ B_{i} ~ sin(i \theta)
\end{equation}
\end{center}
here $A_{0}$ is the zero point, $\theta$ is the phase, while $A_{i}$ and $B_{i}$ are the amplitude parameters \citep{Sca82}. The synthetic light curves derived by Equation (1) were assumed as the quiescent level for each flare. Two examples for the flares detected from the observations and their quiescent levels are shown in Figure 6. After modelling light variations out of both the eclipses and flares, the flare parameters are calculated, as it was previously described by \citet{Dal12} in detail.

The flare rise ($T_{r}$) and decay ($T_{d}$) times, the flare amplitudes and the equivalent durations ($P$) were calculated after the start and end points of a flare were determined. These flare parameters calculated for each flare detected from entire data sets are listed in Table 3. In this table, the times of flare maxima in the first column, the flare equivalent durations (s), the flare rise times (s), the flare decay times (s), the flare total times (s) and the flare amplitudes in the last column were listed, respectively.

225 flares in the total were detected from the available observational data of the system taken from the Kepler Database. The equivalent duration values for each flare are calculated by Equation (2) given by \citet{Ger72}.

\begin{center}
\begin{equation}
P = \int[(I_{flare}-I_{0})/I_{0}]dt
\end{equation}
\end{center}
where $P$ is the flare equivalent duration in seconds, $I_{flare}$  is flux at the moment of a flare, and $I_{0}$ is the flux of the system in the quiescent level, which were modelled by the Fourier method for the parts out-of-eclipses. Considering the reason explained by \citet{Dal10, Dal11}, which is mentioned in Section 1, the flare energies were not used in this study. In order to derive the models, the equivalent duration parameter was used instead of the flare energy. Using the orbital period of the system, the phases for each flare was calculated, and the phase distribution of these 225 flares is shown in Figure 7. In the figure, it is plotted the flare total number computed for each phase range of 0.10.

When the relationship between some flare parameters is examined, it can be seen that the flare equivalent duration varies versus the flare total time within a certain rule. In fact, as it had been demonstrated by \citet{Dal10, Dal11}, the regression calculations processed with the SPSS V17.0 \citep{Gre99} and GraphPad Prism V5.02 \citep{Daw04} programs in this study show that the One-Phase Exponential Association (hereafter OPEA) is the best function to model the distribution of the flare equivalent durations in the logarithmic scale versus the flare total time. The OPEA function is a special function that has a term "$Plateau$" and is represented by Equation (3) \citep{Mot07, Spa87}:

\begin{center}
\begin{equation}
y~=~y_{0}~+~(Plateau~-~y_{0})~\times~(1~-~e^{-k~\times~x})
\end{equation}
\end{center}
where the parameter $y$ is equivalent duration in the logarithmic scale, $x$ is the total time of a flare, and $y_{0}$ is the theoretical equivalent duration in the logarithmic scale for the minimum flare total time, as it had been previously defined by \citet{Dal10}. In other words, the parameter $y_{0}$ defines the minimum equivalent duration that someone can obtain for a flare occurring on the active component. Therefore, $y_{0}$ value depends on the brightness of the observed target and on the sensitivity of the used optical system. The $Plateau$ value defines the upper limit of the equivalent durations for the flares observed on a particular star. According to \citet{Dal11}, this parameter is defined as the saturation level for the flare activity in the observed wavelength range for the observed target.

Using the least-squares method, the distribution of the equivalent duration versus the flare total time was modelled by the OPEA function for the flares. The obtained model is shown in Figure 8 with a 95$\%$ confidence interval. The computed parameters from the OPEA model are listed in Table 4. The span value listed in the table is the difference between Plateau and $y_{0}$ values. The $half-life$ value is half of the first $x$ value, which the flare equivalent durations in the logarithmic scale reach the maximum level defined as the $Plateau$ value. In other words, it is half value of the first total time, where the highest flare energy is seen for the flare.

In the Kepler Mission database, KOI-256's short cadence data are available for the time range from BJD 24 55372.460219 to BJD 24 55552.55836 and from BJD 24 56419.80351 to BJD 24 56424.01160. In total, the KOI-256 has been observed \textbf{in short cadence format} during 184.30624 days (4423.34976 hours). From these data, 225 flares were obtained and their total equivalent durations were computed of 8169.834 seconds. In the literature, two different flare frequencies were defined by \citet{Ger72}. These frequencies are given by Equation (4, 5):

\begin{center}
\begin{equation}
N_{1}~=~\Sigma n_{f}~/~\Sigma T_{t}
\end{equation}
\end{center}

\begin{center}
\begin{equation}
N_{2}~=~\Sigma P~/~\Sigma T_{t}
\end{equation}
\end{center}
where $\Sigma n_{f}$ is the total number of the obtained flare, while $\Sigma T_{t}$  defines the total observing time of the star. $\Sigma P_{u}$ is the sum of the equivalent durations of all detected flares. According to these definitions, the $N_{1}$ frequency was found to be $0.05087$ $h^{-1}$, while the $N_{2}$ frequency was found to be $0.00051$.

\citet{Ger72} described the flare frequency distribution separately calculated for the different energy limits of the flares detected from a star, which defines the flare energy character for that star. However, using the flare equivalent duration instead of the flare energy parameter in this study, since the flare energy parameter depends on the quiet intensity level of the star in the observing band, the flare frequencies were calculated for different flare equivalent duration limits for the 225 flares. The obtained cumulative flare frequency distribution is shown in Figure 10.

\subsection{Orbital Period Variation}

The minima times were computed from the KOI-256's short cadence data from the first quarter to the Quarter 17, which were taken from the publicly available Kepler Database, without any corrections. The minima times were computed with a script depending on the method described by \citet{Kwe56}. In this method, taking symmetrically increasing and decreasing parts of minima, the minima times were computed by fitting this parts with polynomial function. Before computing the minima times, the flares as the activity exhibited by the system were removed from the light curves. In the second step, $(O-C) I$ residuals were determined for the obtained minima times. However, some of them include very large errors. It is seen that the flare activity occurred during these minima, when the light curves were examined for these minima times with large errors. Thus, these minima times were removed from the $(O-C)$ data. As a result, 125 minima times were determined in the analyses. The obtained $(O-C) I$ residuals were adjusted by the linear correction given in Equation (6):

\begin{center}
\begin{equation}
JD~(Hel.)~=~24~54965.55513 (4)~+~1^{d}.3786503(1)~\times~E.
\end{equation}
\end{center}
The computed minima times $(O-C) I$ and the $(O-C) II$ residuals obtained by applying a linear correction are listed in Table 5. In the table, the minima times, cycles, the minima type, $(O-C) I$ and $(O-C) II$ residuals are listed, respectively. An interesting variation is seen in the variation of obtained $(O-C) II$ residuals versus time in Figure 11.

Firstly, checking the minima types given in the third column of Table 5, it was practically examined whether there is a marker about the existence of secondary minimum, which is the subject of discussion in the literature. Secondly, it was examined whether there was any separation in the $(O-C) II$ residuals plotted in Figure 11. If there was any secondary minima, it would be expected a separation between the primary and secondary minima time residuals due to stellar spot activity, as it was demonstrated by \citet{Tra13} and \citet{Bal15} for the first time. However, as it is seen from Figure 11, there is no decomposition in the minima time residuals as the primary and secondary minima.

\section{Results and Discussion}

The analysis of KOI-256's data taken from the Kepler Database \citep{Sla11, Mat12} indicates that the system has high chromospheric activity. However, it is necessary to compare the activity level between similar stars in order to reach a more definite result about the activity nature. In the literature, KOI-256 is an eclipsing binary system, with a white dwarf as a primary component and a main sequence star from M3 spectral type as a secondary component \citep{Mui14}. The temperatures of the components were determined as $T_{1}=$7100 K and $T_{2}=$3450 K in the analyses of the spectral data together with the photometrical data of the system \citep{Mui14}.

The eclipsing binary system KOI-256 was observed in \textbf{the short cadence format} from HJD 24 54964.51238 to HJD 24 56424.011602 in total of 184.30624 days (4423.34976 hours). Within this study, 225 flares were determined and the parameters of each flare were calculated. Using the frequency descriptions defined by \citet{Ger72}, the flare frequencies for KOI-256 were calculated as $N_{1}=$0.05087 $h^{-1}$ and $N_{2}=$0.00051, respectively. The flare frequencies of KOI-256 were compared with the flare frequencies found from the young main sequence dMe dwarfs, known as UV Ceti-type, exhibiting flare activity. The flare frequencies of KIC 9761199 were found to be $N_{1}=$0.01351 $h^{-1}$ and $N_{2}=$0.00006 \citep{Yol16}, while they were found to be $N_{1}=$0.01735 $h^{-1}$ and $N_{2}=$0.00001 for KIC 9641031 \citep{Yol17}. KOI-256, when compared to these two systems, seems to have the highest flare frequency values among similar systems. On the other hand, the $N_{2}$ frequencies were computed to be 0.088 for EQ Peg and 0.086 in the case of AD Leo, while the $N_{1}$ frequencies were determined as 1.331 $h^{-1}$ for AD Leo, a UV Ceti type single star, and 1.056 $h^{-1}$ for EV Lac \citep{Dal11}. Compared to these single stars of the UV Ceti type, the flare frequencies of KOI-256 are found to be quite low. However, it is known in the literature that the flare frequency of EV Lac varies with time \citep{Let97}. As it is seen in Figure 9, the monthly frequencies vary with time like EV Lac, when the flare frequencies are computed for each month in the case of KOI-256.

The $Plateau$ value of the OPEA model depending on the flare equivalent duration distribution in the logarithmic scale versus the flare total time is found to be 2.3121$\pm$0.0964 s over detected 225 flares. This value is 3.014 s for EV Lac ($B-V=1^{m}.554$), 2.935 s for EQ Peg ($B-V=1^{m}.574$) and 2.637 s for V1005 Ori ($B-V=1^{m}.307$) \citep{Dal11}. The $Plateau$ value was found to be 1.232$\pm$0.069 s for KIC 9641031 ($B-V=0^{m}.74$) \citep{Yol16} and 1.951$\pm$0.069 for KIC 9761199 ($B-V=1^{m}.303$) \citep{Yol17}. If it is considered that one of the components is a white dwarf in the case of KOI-256, it is clear in this study that the chromospherical active component is the main sequence star from the spectral type of M3. The B-V color index for this component is given as ($1^{m}.42$) in the literature \citep{Wal13b}. Therefore, the flares with high energy exhibited by KOI-256 are strong enough to be compared with the flares exhibited by the single UV Ceti type stars. \citet{Dal11}. found that the $Plateau$ value is constant for a star, but varies from one star to another depending on their $B-V$ color indexes. The authors have defined this value as the saturation level of flares on a star, which is maximum energy limit a flare can reach.

On the other hand, the $half-life$ parameter of the OPEA model was found to be 2233.6 s for KOI-256. This value is almost 10 times higher than the value obtained on single flare stars. This value is 433.10 s for DO Cep ($B-V=1^{m}.604$), 334.30 s for EQ Peg and 226.30 s for V1005 Ori \citep{Dal11}. In the same way, the $half-life$ parameter is 2291.7 s for KIC 9641031, while it is 1014 s for KIC 9761199, which are binary systems including a dMe type component \citep{Yol16, Yol17}. As a result, when the flare total time for the stars such as EQ Peg, V1005 Ori and DO Cep reaches a few minutes, the their flare energies can easily reach the Plateau level. However, in the case of KOI-256, a flare event must last at least 37 minutes, about a few tens of minutes, in order to reach the maximum flare energy described as the Plateau level for this system. Similar durations are also observed in the case of both flare rise and total times. The maximum flare rise time ($T_{r}$) observed in the single UV Ceti type stars is about 2042 s for V1005 Ori, 1967 s for CR Dra \citep{Dal11}, while the longest flare rise time obtained in KOI-256 is 3942.749 s. The longest flare rise time obtained flares on KIC 9641031 is 5178.87 s \citep{Yol16}, while it is 1118.099 s for KIC 9761199 \citep{Yol17}. Similarly, the maximum flare total time for the flares obtained on the V1005 Ori is 5236 s, while it is 4955 s for CR Dra flares \citep{Dal11}. However, the longest flare obtained on KOI-256 lasts along 22185.361 s.

Considering the stellar spot activity together with the flare activity detected from KOI-256, it is obvious that the system exhibits a high level of chromospheric activity. Using very sensitive observations provided by Kepler Mission \citep{Jen10a, Jen10b}, some variations with small amplitudes, which are impossible to observe with ground-based classic telescopes, can be easily detected. When the system light variations out-of-eclipses without any flares is carefully examined, it is seen that the shape of eliminated light curve can vary per two or three cycles at most. This variation is caused due to both rapid evolutions and migrations of the cool stellar spots on the surface of the M3 star. It needs about 4-5 days for the evolution of the stellar spots on the active component in the case of KOI-256, while it takes a few weeks for the cool spot groups on the solar surface \citep{Ger05}. In this study, the stellar spot distribution on the surface of the active component was modelled by the SPOTMODEL program \citep{Rib02, Rib03}, though there are some deficiency in the method of this program. As it was mentioned in Section 1, the method using in the SPOTMODEL program do consider neither the spot evolution nor spot migration on the surface during long time. However, separating the data into suitable sub-sets by considering the light curve shape change point, we solved out this problem. The whole data were separated into 138 sub-sets. The cycle shapes of sinusoidal light variation are different from one sub-set data to the next, while the shapes of the cycles are the same among themselves in each sub-set data. Thus, each sub-set data was individually modelled. On the other hand, maximum level of the obtained brightness is not suitable for modelling by the SPOTMODEL program due to detrended data. However, the spot longitude parameters is calculated in this study, because it is an important parameter to reveal the spot migration instead of the spot latitude or radius.

The result parameters of the modelling spot variation are plotted in Figures 2, 3 and 4. As it is seen from Figure 2, the models of 138 sub-sets indicate that two cool spots (or spotted areas) are enough to absolutely fit the observed light variation out-of-eclipses. Although both latitude and radius parameters of the spots do not externalize the realistic nature of the spots on the stellar surface as discussed earlier, their variations are also plotted in Figure 3 to note as an initial approach. The main goal of these models is longitudinal variation and it is shown in Figure 4. Observations of the cool spot groups on solar surface reveal the existence of two permanently active longitudes separated by 180 degree from each other. These active longitudes known as the Carrington Coordinates are constant structure according to some authors, while some authors indicate that the rotation speeds of these active longitudes are not constant, but slowly change \citep{Lop61, Sta72, Bog82}. Similarly, in this study, it was found that there are two stellar cool spots on the active component of KOI-256, which show the migration behaviour with different speeds. Modelling the migration movements shown in Figure 4 by the linear fits, it is estimated that the migration of the first stellar spot is 3.95 years and the migration period of the second one is 8.37 years.

To test whether these results are real, it was tested by another method. Computing $\theta_{min}$ parameter, which is the phases of the observed sinusoidal variation out-of-eclipses, the migration behaviour of the dominant spotted area was tried to find out. Indeed, as it is seen from Figure 5, the $\theta_{min}$ values obtained for all cycles were plotted versus Heliocentric Julian Day, and then, its trend was fitted by a linear function. The migration period of dominant spotted area is found to be 9.13 year. In addition, the most impressive finding was came from the residual $\theta_{min}$ according to the linear fit. The residuals show the same trend with the residuals of the longitudinal variation obtained from the SPOTMODEL analyses. This situation absolutely reveals that the main goal of the SPOTMODEL analyses have been obtained, and it does figure out the migration of active regions close to its real nature.

However, as can be seen in Figure 4, some sinusoidal variations still remain, after a linear correction is applied to the migration movement. All of these migration behaviours are an indicator of the strong differential rotation on stellar surface. However, a high-resolution spectroscopic observation is needed to confirm this case. In the literature, the system's age is given as 0.01 Gyr by \citet{Wal13b}. However, as it is described in Section 1, using the Equation 2.3 given by \citet{Gan97} with the stellar parameters taken from \citet{Mui13},the age of the system is estimated as $\sim$2 Gyr in this study. Considering the law expressed by \citet{Sku72}, this age is a bit old for rapid rotation, considering high level chromospheric activity. However, KOI-256 is a binary system. In this case, tidal effects can let the components to rotate rapidly, which lets the system has a significant influence on both the flare and spots activity behaviour.

\citet{Mui14} listed the semi-major axis length of the system as $a= 4.51$ $R_{\odot}$, giving the radii of components as $R_{1}=$0.54 $R_{\odot}$ and $R_{2}=$0.01 $R_{\odot}$. This situation indicates that the components are very close to each other, which can cause rapid activity variations due to some tidal effects on each other. On the other hand, the most interesting result in term of stellar spot and flare activity exhibited by the system is that the stellar spot areas exhibit a rapid variation in location due to the migration movement as shown in Figure 5, while it is seen interestingly that the flares are in aim to occur frequently between the phases of 0.10 - 0.20 and also 0.60 - 0.70, when the averaged phases are calculated in each 0.10 phase range for 225 flares. In the case of both spot and flare activities, the phases were computed with using the orbital period of binary system with a white dwarf and a main sequence M3 components. This case is an unexpected situation, because the locations of flares are quite stable compared to the behaviour of the spot activity. Although flare activity is seen in all phases, it is observed that the flare activity most frequently occurs in these phase intervals. The location of the white dwarf is as close as 4.51 $R_{\odot}$ to the active component, which indicates that the components are interacting magnetically. A possible explanation is that the quickly evolving stellar spot areas under both the differential ration and tidal effects exhibit rapid migration movements, while the flares are frequently occurring in some definite longitudes on the active component surface due to the magnetic interaction of the white dwarf and active components. This situation leads KOI-256 to take an important position among its analogues for especially the future spectral studies.

125 minima times were determined from the all available data. It is seen that the light curve has only the primary minimum, when entire the light curves are examined. According to \citet{Tra13} and \citet{Bal15}, the $(O-C) II$ residuals of the primary and secondary minima show some sinusoidal variation versus time, but in the opposed phase, if one of the components in an eclipsing binary system is a magnetically active star. However, the $(O-C) II$ residuals obtained from the minima detected from the light curves of KOI-256 do not exhibit any separation. This situation also confirms that all the minima detected from the light curves are primary ones.

As it is seen from Figure 11, considering all 125 minima times, there is not any clear variation in the $(O-C) II$ residuals distribution versus time obtained by improving linear correction. As shown in Figure 11, the $(O-C) II$ residuals scatter in a range of about 50-60 sec. Whereas according to \citet{Sza03} in the literature, KOI-256 shows an $(O-C) II$ variation with an amplitude of 0.4464 minute and a period of 41.755397 day. However, the results obtained in this study do not confirm the variation found by \citet{Sza03}. \textbf{The problem, which cause the different results can be due to the method used to compute the minima times or due to the different data formats used in these studies. In this study, the minima times were computed from the short cadence data, while the minima times were calculated from the long cadence data by \citet{Sza03}.} In addition, although the flares as the activity exhibited by the system were removed from the light curves before computing the minima times, the sinusoidal variation due to the stellar spot activity was not cleared from the data in order to test whether the primary and secondary minima were separated, as described by \citet{Tra13} and \citet{Bal15}. Although there is no any clue for the secondary minima, the situation can cause a scattering in a wide range for the $(O-C) II$ residuals variation obtained in this study.

\section*{Acknowledgments} We wish to thank the Turkish Scientific and Technical Research Council for supporting this work through grant No. 116F213. We also thank the referee for useful comments that have contributed to the improvement of the paper.

\clearpage

\begin{figure*}
\begin{center}
\includegraphics[width=16cm]{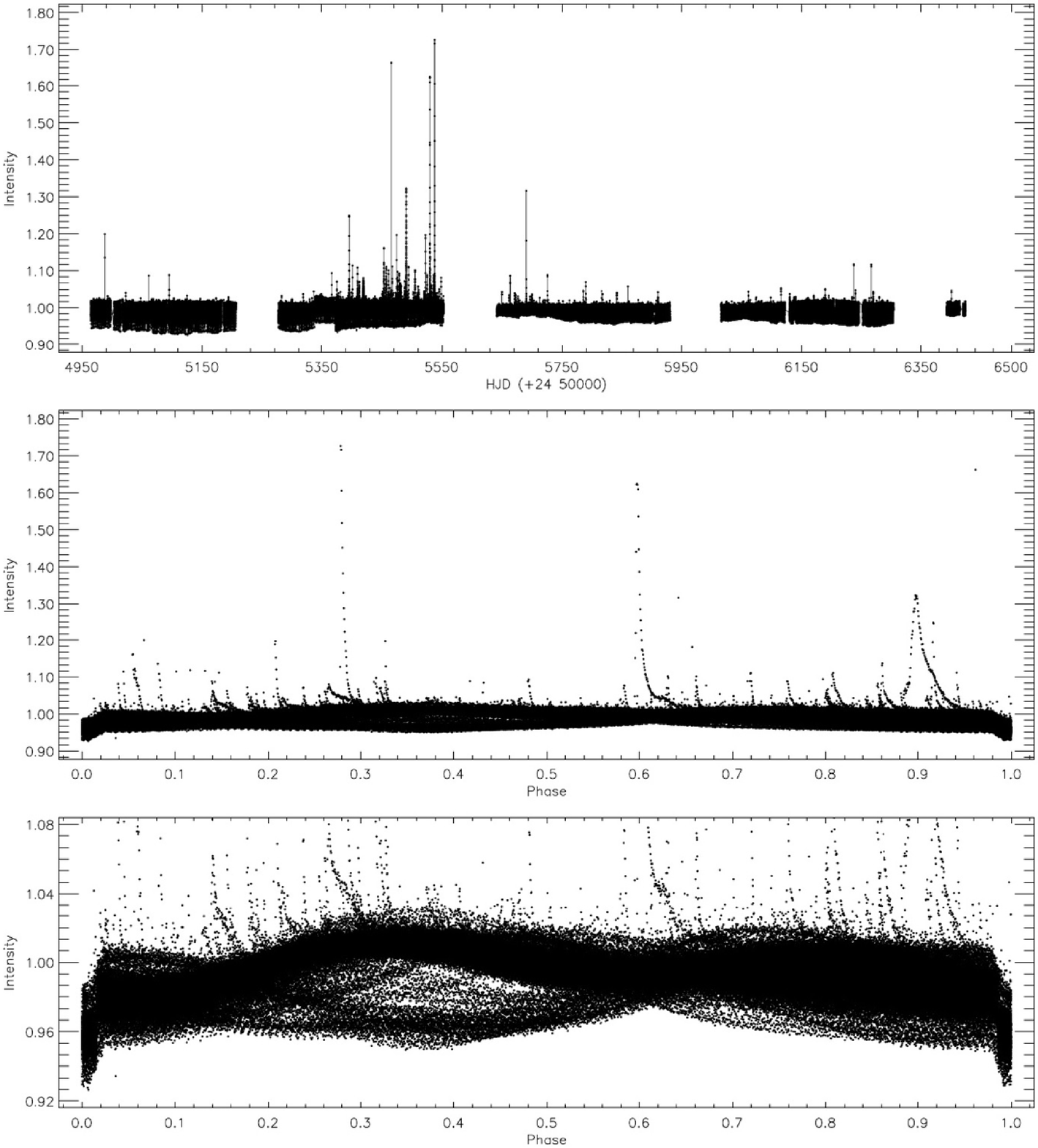}
\vspace{0.7 cm}
\caption{All of the Long Cadence and Short Cadence data taken from the Kepler Database for the KOI-256 are shown. The light variation was plotted in the plane of intensity taken from database as detrended form versus the phase computed by using orbital period.}
\label{Fig. 1.}
\end{center}
\end{figure*}

\begin{figure*}
\begin{center}
\includegraphics[width=15cm]{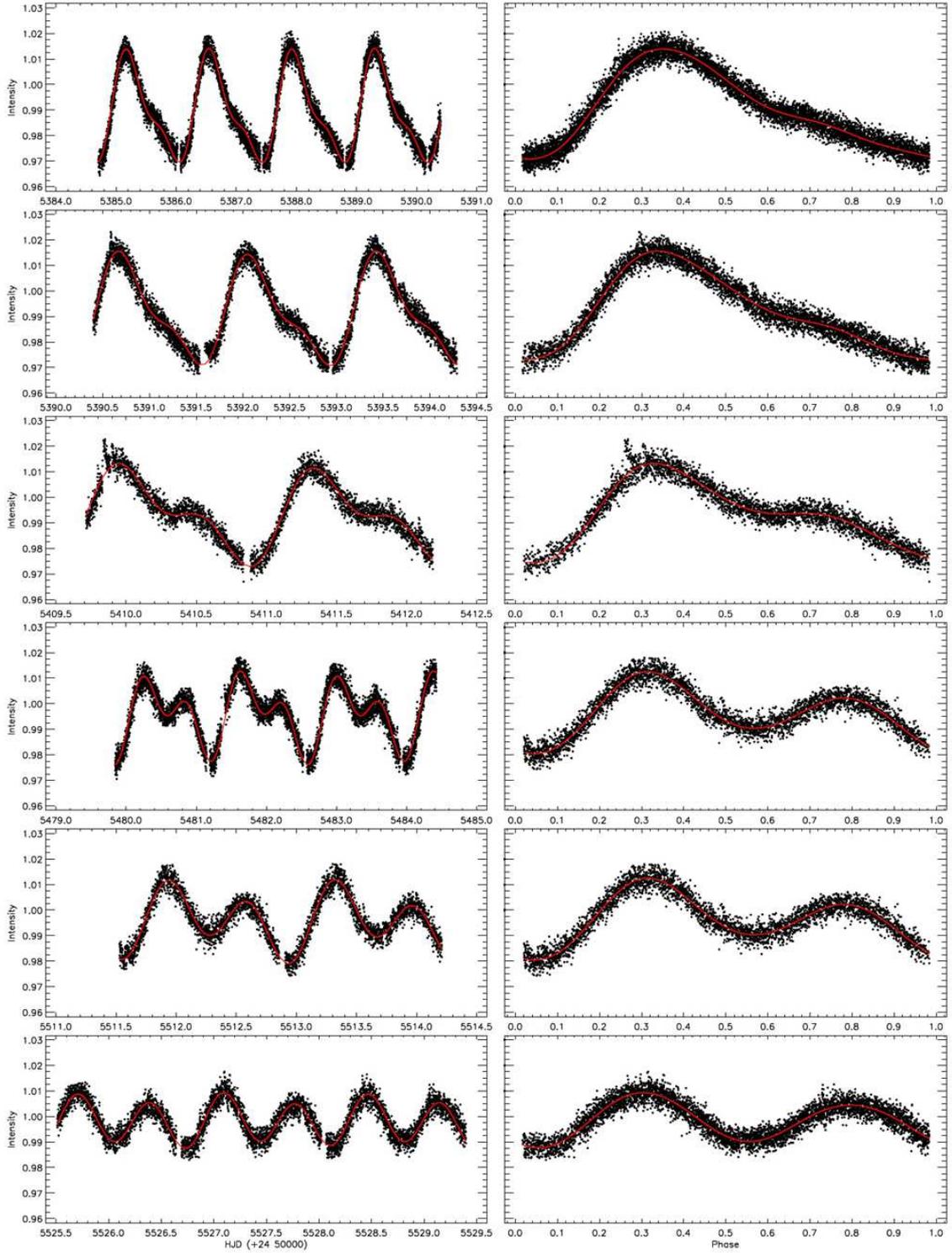}
\vspace{0.7 cm}
\caption{The observed light curve samples and their synthetic model fits derived by the SPOTMODEL analyses. In the figure, the filled circles represent the observations as intensity in detrended form, while the lines (red) represent the synthetic fits. In the left panels, the data and synthetic model are plotted versus time as  Heliocentric Julian Day, while they are plotted versus phase computed using epoch and orbital period given in Equation (6) in the right panels.}
\label{Fig. 2.}
\end{center}
\end{figure*}

\begin{figure*}
\begin{center}
\includegraphics[width=20cm]{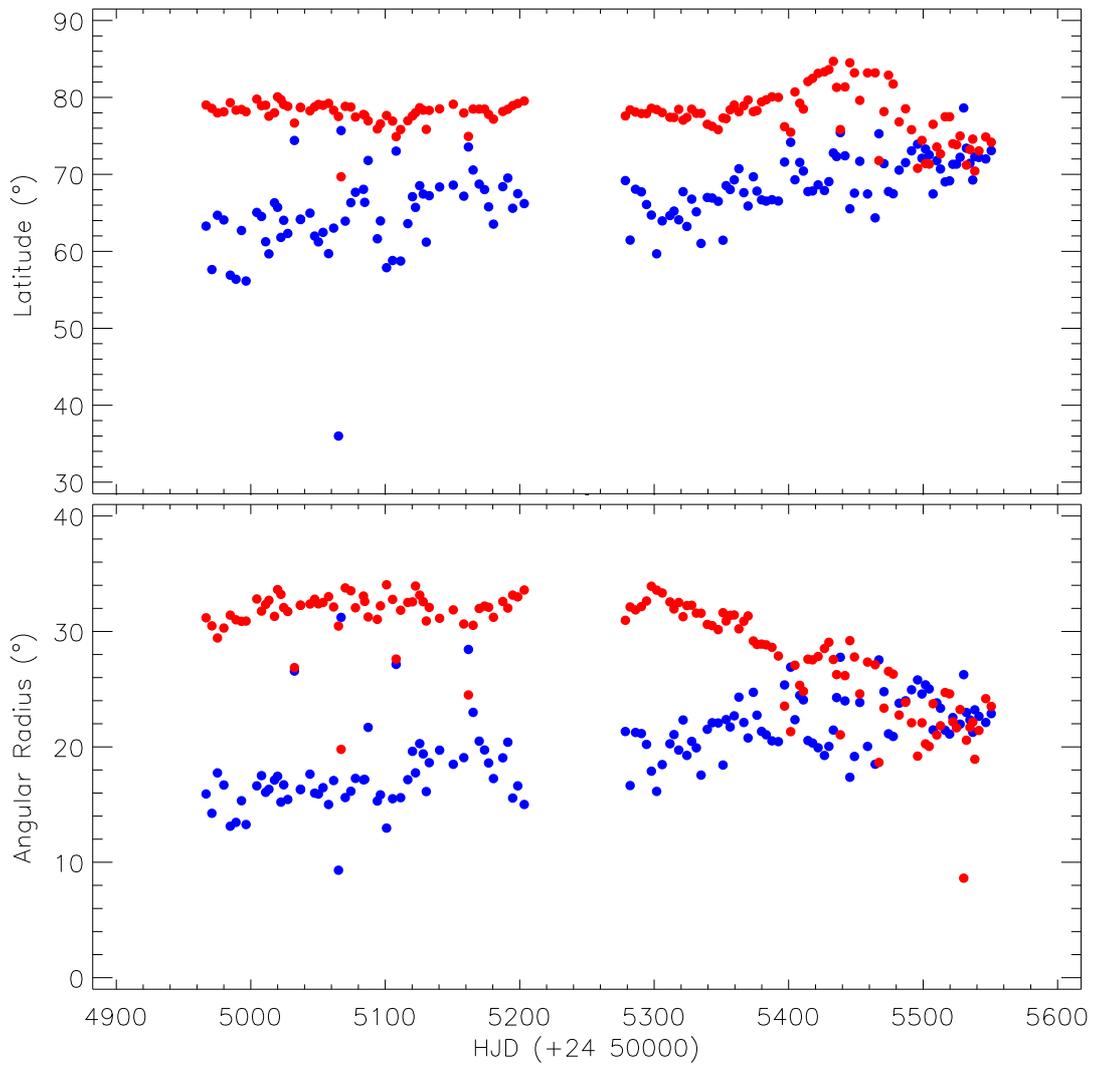}
\vspace{0.7 cm}
\caption{The variations of both spot latitude and spot radius parameters obtained with the SPOTMODEL program versus time. The filled blue circles represent the first spot and the filled red circles represent the second spot.}
\label{Fig. 3.}
\end{center}
\end{figure*}

\begin{figure*}
\begin{center}
\includegraphics[width=16cm]{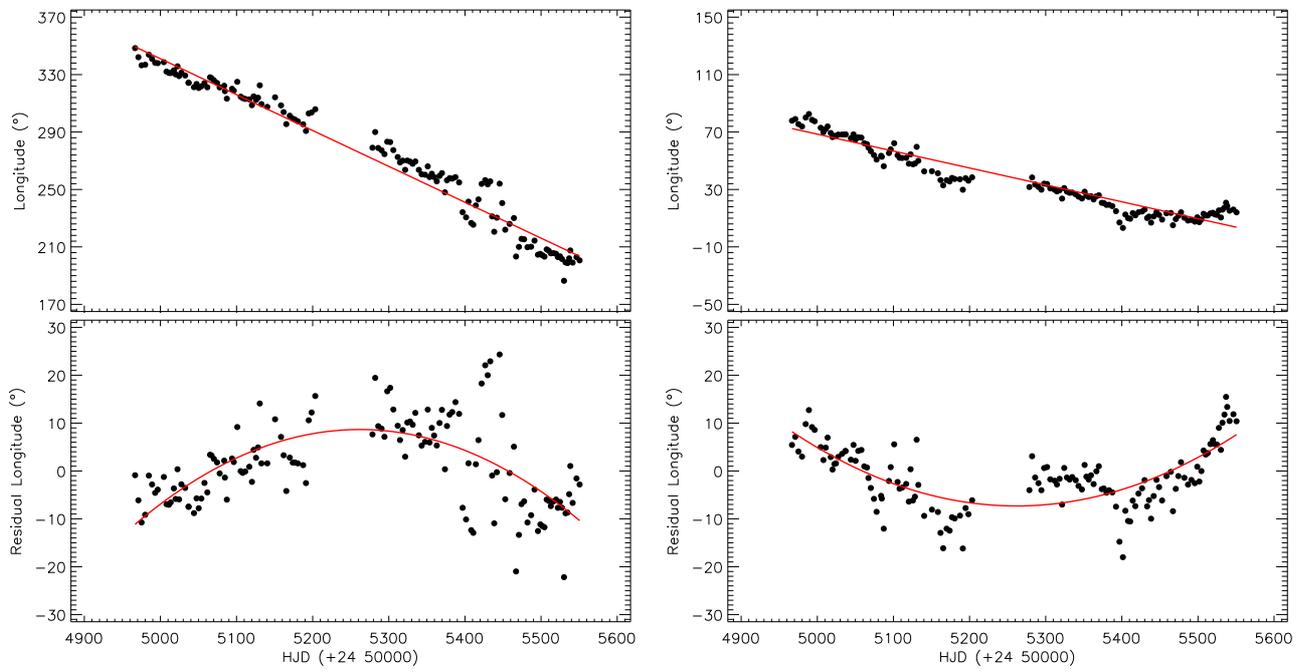}
\vspace{0.7 cm}
\caption{The variations of the first spot longitude as degree (left) and the second spot (right) are shown in the upper, while the residuals after the linear correction to the longitude variations are shown in the lower panel.}
\label{Fig. 4.}
\end{center}
\end{figure*}

\begin{figure*}
\begin{center}
\includegraphics[width=18cm]{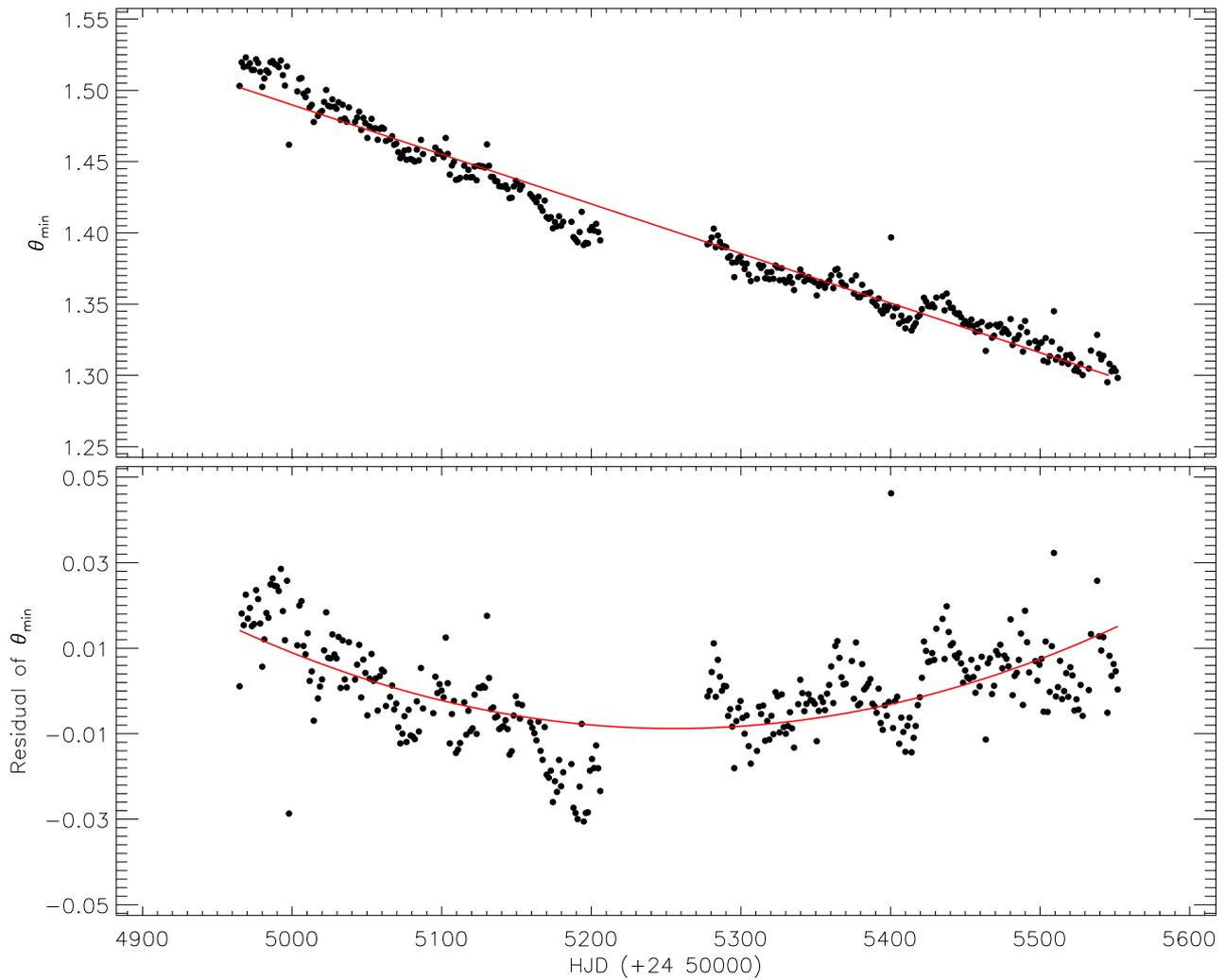}
\vspace{0.7 cm}
\caption{$O-C$ diagram for the observed minima times of sinusoidal variation versus time in each data subset shown with $\theta_{min}$ term and its linear fit are shown in the upper panel. The filled circles represent the $\theta_{min}$ variation, while the line (red) show the linear fit. In the bottom panel, the residuals of $\theta_{min}$ is shown with its parabolic fit, which is plotted to show the trends clearly for the readers.}
\label{Fig. 5.}
\end{center}
\end{figure*}

\begin{figure*}
\begin{center}
\includegraphics[width=15cm]{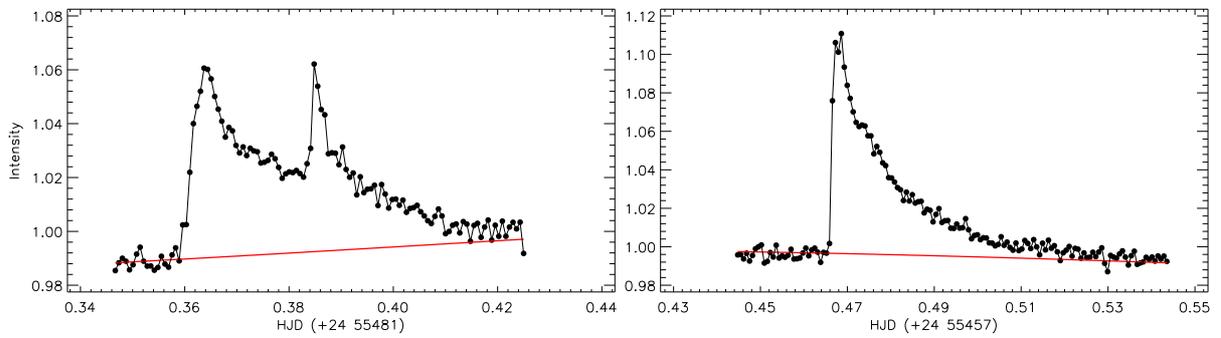}
\vspace{0.7 cm}
\caption{Two flare examples detected from the system. The light variation was plotted in the plane of intensity taken from database as detrended form versus time. The filled black circles represent the observations, while the red lines represent the quiescent level modelled by the Fourier method.}
\label{Fig. 6.}
\end{center}
\end{figure*}

\begin{figure*}
\begin{center}
\includegraphics[width=16cm]{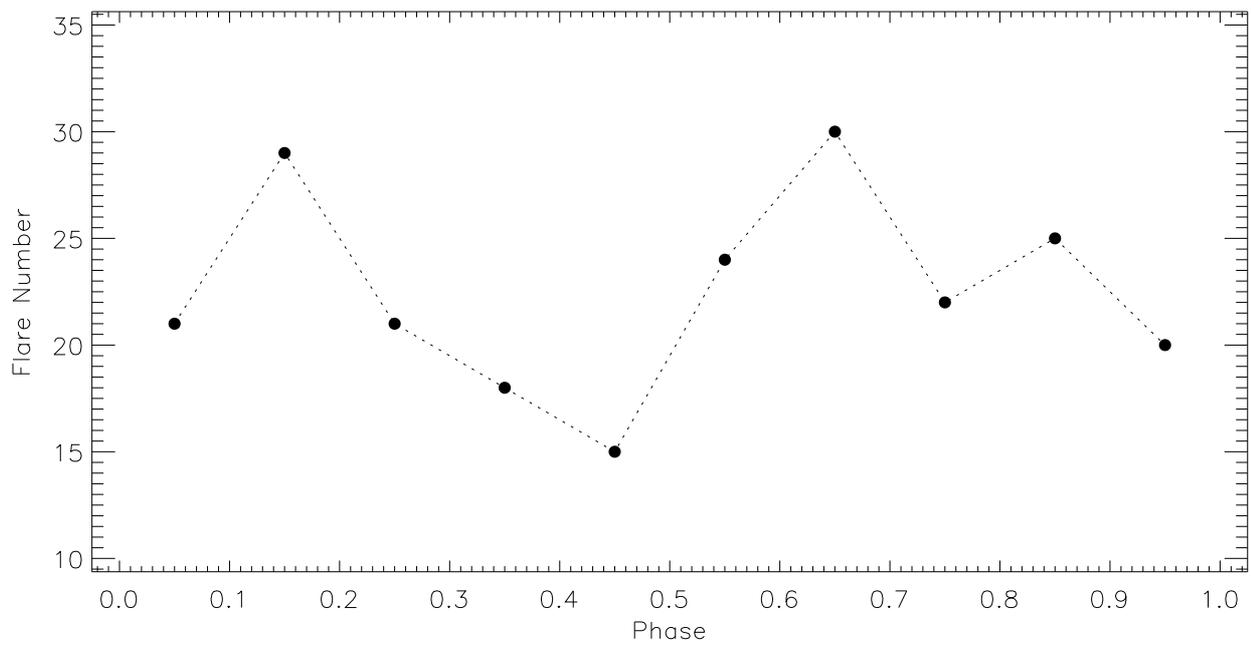}
\vspace{0.7 cm}
\caption{The distribution of flare total number in phase range of 0.10 is plotted versus the phase computed by using orbital period for 225 flares.}
\label{Fig. 7.}
\end{center}
\end{figure*}

\begin{figure*}
\begin{center}
\includegraphics[width=16cm]{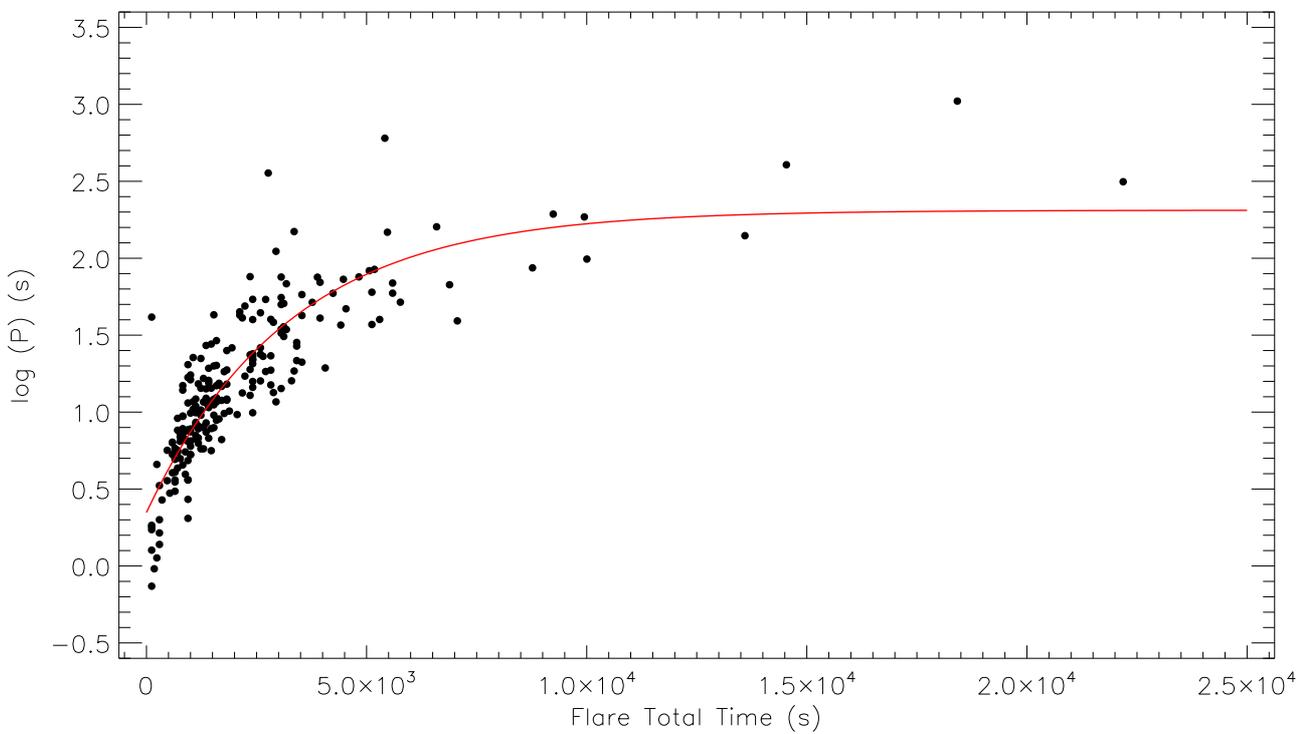}
\vspace{0.7 cm}
\caption{The distribution of the equivalent duration in the logarithmic scale ($log P$) are plotted versus the flare total time, which were sum of flare rise and decay times. The OPEA model obtained over 225 flare determined in the analyses. The fill circles represent the observed flares, while the red line represents the OPEA model.}
\label{Fig. 8.}
\end{center}
\end{figure*}

\begin{figure*}
\begin{center}
\includegraphics[width=20cm]{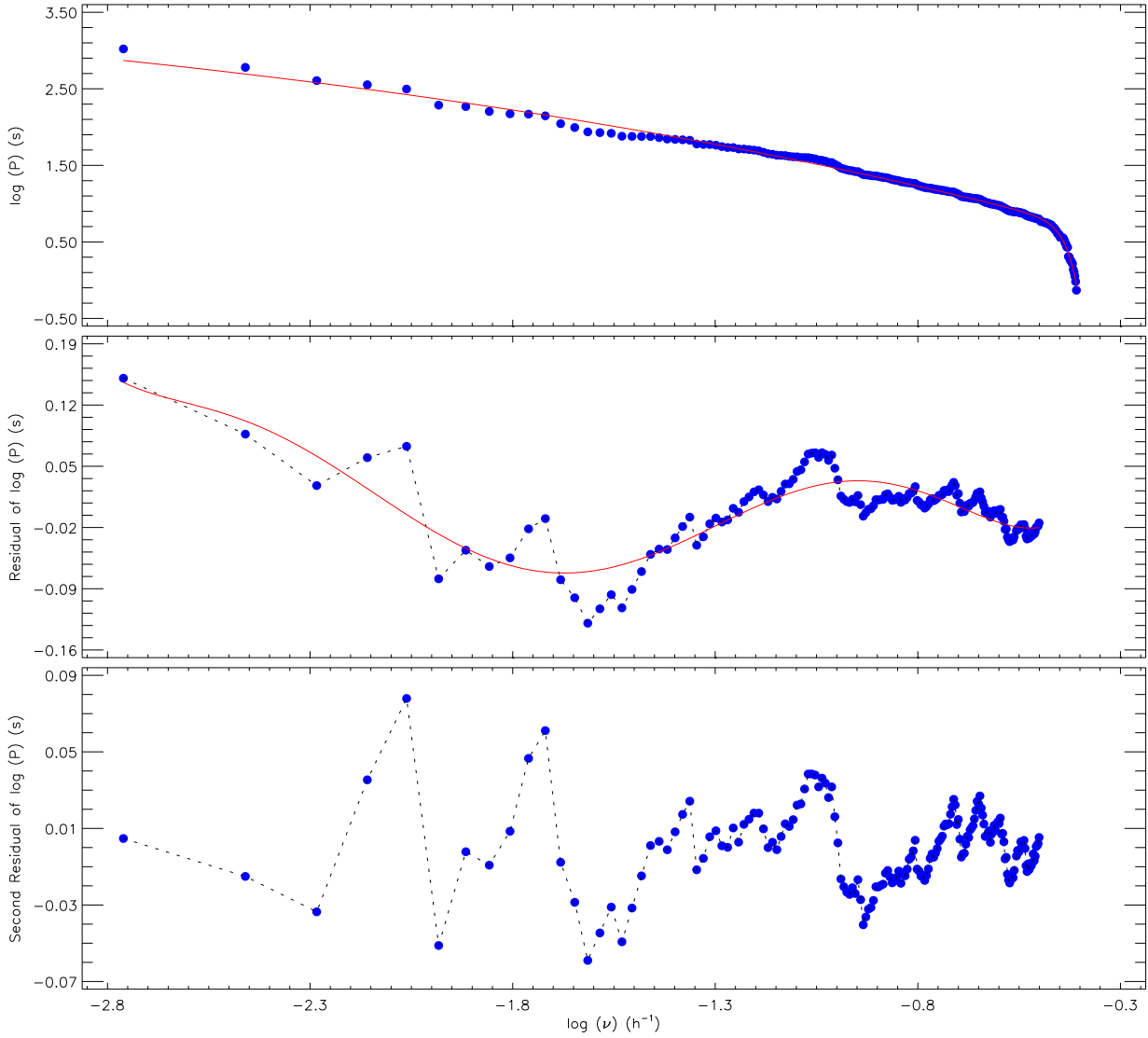}
\vspace{0.7 cm}
\caption{The cumulative flare frequencies ($log \nu$) and model computed for 225 flares obtained from KOI-256. In the upper panel, it is seen the variation of the flare equivalent durations in logarithmic scale ($log P$) versus the cumulative flare frequency, which is called the flare energy spectrum \citep{Ger05}, while the residuals obtained from the model are shown in the middle and bottom panels.}
\label{Fig. 9.}
\end{center}
\end{figure*}

\begin{figure*}
\begin{center}
\includegraphics[width=16cm]{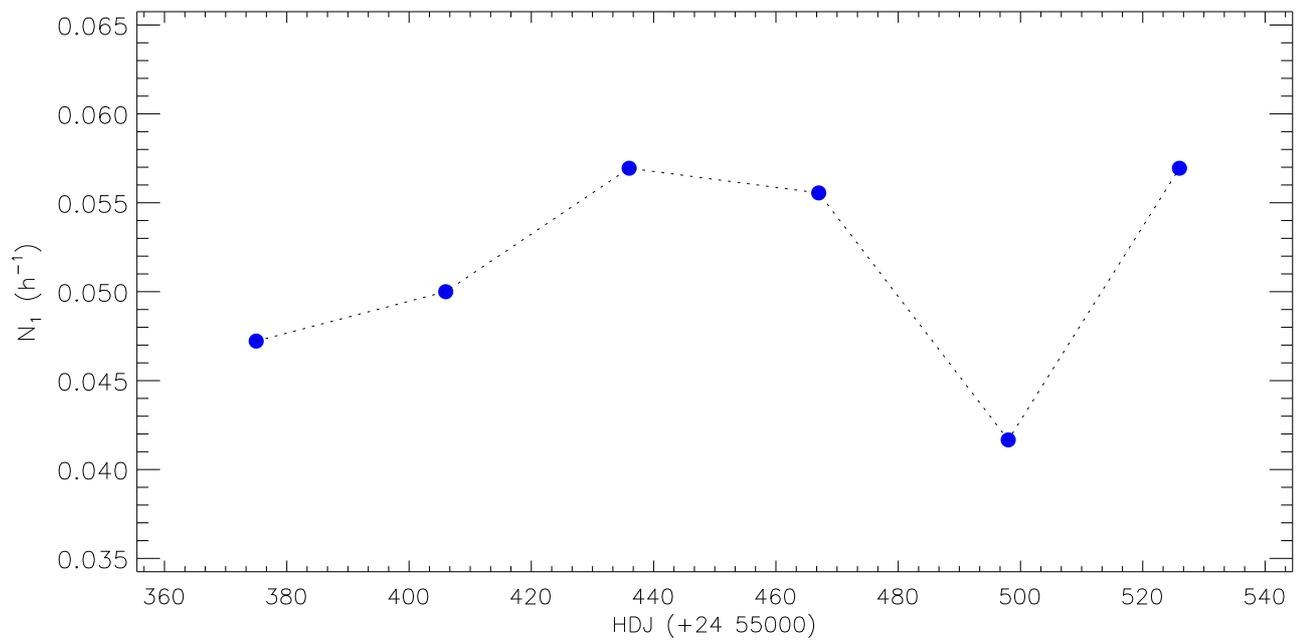}
\vspace{0.7 cm}
\caption{The monthly variation of the flare frequency of $N_{1}$, which indicates total flare number per each hour, for KOI-256 is shown for the entire observing season.}
\label{Fig. 10.}
\end{center}
\end{figure*}

\begin{figure*}
\begin{center}
\includegraphics[width=15cm]{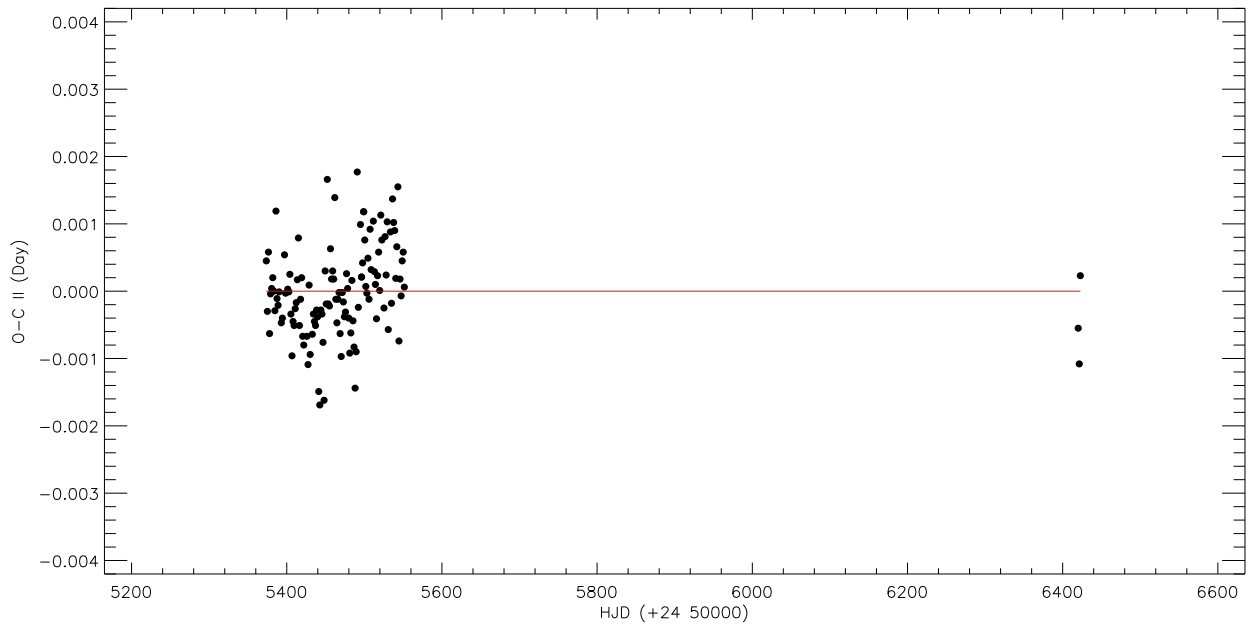}
\vspace{0.7 cm}
\caption{The variations of minima time's residuals $(O-C) II$ obtained by applying a linear correction to minima times are shown versus time. All the residuals are shown with filled circles, while the red line represents a linear fit.}
\label{Fig. 11.}
\end{center}
\end{figure*}

\clearpage

\begin{table*}
\begin{center}
\caption{Physical parameters of KOI-256 taken from the literature.}
\begin{tabular}{lcr}
\hline
\hline
$P_{orb}$ &	($day$)	& 1.3786548 $(^1)$\\
$a$	& ($AU$)	& 0.0250 $(^1)$\\
$i$	& ($^\circ$)	& 89.3 $(^2)$\\
$(B-V)$	& ($^m$)	& 1.42 $(^3)$\\
$R_{1}$	& ($R_{\odot}$)	& 0.540 $(^1)$\\
$R_{2}$	& ($R_{\odot}$)	& 0.01345 $(^1)$\\
$T_{1}$	& ($K$)	& 3450 $(^1)$\\
$T_{2}$	& ($K$)	& 7100 $(^1)$\\
$M_{1}$	& ($M_{\odot}$)	& 0.51 $(^1)$\\
$M_{2}$	& ($M_{\odot}$)	& 0.592 $(^1)$\\
$ [M/H]$ &	- & 0.31 $(^1)$\\
$d$	& ($pc$) & 155 $(^4)$\\
J, H, K	& ($^m$) & 12.701 - 12.000 - 11.782 $(^5)$\\
Spectral Type &	- & M3 V + WD $(^4)$\\
\hline
\end{tabular}
\end{center}
$(^1)$ - \citet{Mui13} \\
$(^2)$ - \citet{Sla11} \\
$(^3)$ - \citet{Wal13b} \\
$(^4)$ - \citet{Mui14} \\
$(^5)$ - \citet{Zac04} \\
\end{table*}

\setcounter{table}{1}
\begin{table*}
\setlength{\tabcolsep}{0.2 pt}
\begin{center}
\caption{The Spot parameters obtained with SPOTMODEL.}
\begin{tabular}{ccccccc}
\hline
\hline
JD	&	$b_{1}$ (Latitude)	&	$b_{2}$ (Latitude)	&	$g_{1}$ (Radius)	&	$g_{2}$  (Radius)	&	$l_{1}$  (Longitude)	&	$l_{2}$   (Longitude)	\\
\hline		 	 	 						 	 	 						 	 	 					
54966.749927	&	63.277	$\pm$	3.430	&	79.015	$\pm$	0.252	&	15.917	$\pm$	1.808	&	31.179	$\pm$	0.410	&	77.848	$\pm$	1.494	&	378.418	$\pm$	2.961	\\
54971.061541	&	57.628	$\pm$	4.704	&	78.572	$\pm$	0.234	&	14.244	$\pm$	1.643	&	30.481	$\pm$	0.313	&	79.038	$\pm$	1.083	&	342.094	$\pm$	2.181	\\
54975.230111	&	64.687	$\pm$	2.303	&	77.990	$\pm$	0.323	&	17.743	$\pm$	1.498	&	29.433	$\pm$	0.492	&	75.469	$\pm$	1.578	&	336.431	$\pm$	3.112	\\
54979.960614	&	64.074	$\pm$	2.669	&	78.130	$\pm$	0.250	&	16.693	$\pm$	1.533	&	30.294	$\pm$	0.401	&	73.857	$\pm$	1.447	&	336.859	$\pm$	2.638	\\
54984.742199	&	56.901	$\pm$	5.588	&	79.318	$\pm$	0.190	&	13.122	$\pm$	1.723	&	31.410	$\pm$	0.259	&	80.089	$\pm$	1.018	&	343.844	$\pm$	1.955	\\
54988.992489	&	56.368	$\pm$	4.173	&	78.341	$\pm$	0.173	&	13.458	$\pm$	1.272	&	31.008	$\pm$	0.204	&	82.514	$\pm$	0.806	&	340.931	$\pm$	1.367	\\
54993.058867	&	62.702	$\pm$	3.014	&	78.436	$\pm$	0.211	&	15.332	$\pm$	1.407	&	30.875	$\pm$	0.289	&	78.478	$\pm$	1.145	&	338.160	$\pm$	1.944	\\
54996.491784	&	56.142	$\pm$	6.666	&	78.138	$\pm$	0.259	&	13.270	$\pm$	2.007	&	30.898	$\pm$	0.325	&	77.494	$\pm$	1.324	&	337.974	$\pm$	2.173	\\
55004.481471	&	65.054	$\pm$	3.030	&	79.779	$\pm$	0.218	&	16.620	$\pm$	1.904	&	32.811	$\pm$	0.426	&	72.950	$\pm$	1.555	&	338.632	$\pm$	3.309	\\
55007.985892	&	64.527	$\pm$	2.464	&	78.929	$\pm$	0.234	&	17.507	$\pm$	1.559	&	31.759	$\pm$	0.409	&	69.841	$\pm$	1.448	&	332.064	$\pm$	2.888	\\
55011.061195	&	61.244	$\pm$	5.076	&	78.986	$\pm$	0.349	&	16.060	$\pm$	2.370	&	32.337	$\pm$	0.489	&	72.068	$\pm$	1.765	&	331.181	$\pm$	3.486	\\
55013.492827	&	59.668	$\pm$	4.704	&	77.560	$\pm$	0.759	&	16.324	$\pm$	1.498	&	32.693	$\pm$	0.289	&	73.878	$\pm$	3.926	&	331.057	$\pm$	2.571	\\
55017.640899	&	66.308	$\pm$	2.303	&	77.999	$\pm$	0.759	&	17.122	$\pm$	1.533	&	31.309	$\pm$	0.325	&	69.324	$\pm$	9.060	&	332.951	$\pm$	3.291	\\
55019.949921	&	65.718	$\pm$	2.988	&	80.074	$\pm$	0.234	&	17.455	$\pm$	2.071	&	33.626	$\pm$	0.482	&	66.448	$\pm$	1.757	&	330.176	$\pm$	3.845	\\
55022.422411	&	61.803	$\pm$	6.938	&	79.715	$\pm$	0.316	&	15.211	$\pm$	3.222	&	33.209	$\pm$	0.581	&	67.316	$\pm$	2.231	&	335.756	$\pm$	4.538	\\
55024.496438	&	64.030	$\pm$	4.381	&	79.073	$\pm$	0.383	&	16.711	$\pm$	2.566	&	32.067	$\pm$	0.613	&	67.204	$\pm$	2.127	&	328.939	$\pm$	4.392	\\
55027.520633	&	62.317	$\pm$	3.422	&	78.843	$\pm$	0.207	&	15.440	$\pm$	1.623	&	31.722	$\pm$	0.325	&	68.190	$\pm$	1.271	&	331.344	$\pm$	2.327	\\
55032.475806	&	74.406	$\pm$	2.185	&	76.669	$\pm$	0.207	&	26.567	$\pm$	3.707	&	26.866	$\pm$	3.598	&	47.323	$\pm$	1.324	&	301.919	$\pm$	2.630	\\
55036.991646	&	64.178	$\pm$	3.007	&	78.693	$\pm$	0.272	&	16.326	$\pm$	1.647	&	32.266	$\pm$	0.363	&	68.294	$\pm$	1.406	&	324.269	$\pm$	2.377	\\
55036.991646	&	64.124	$\pm$	3.020	&	78.697	$\pm$	0.271	&	16.296	$\pm$	1.645	&	32.273	$\pm$	0.361	&	68.314	$\pm$	1.399	&	324.309	$\pm$	2.366	\\
55044.051454	&	64.961	$\pm$	2.458	&	78.258	$\pm$	0.285	&	17.635	$\pm$	1.564	&	32.375	$\pm$	0.403	&	65.672	$\pm$	1.476	&	321.194	$\pm$	2.559	\\
55047.525157	&	61.982	$\pm$	3.970	&	78.775	$\pm$	0.290	&	15.990	$\pm$	1.888	&	32.788	$\pm$	0.375	&	68.351	$\pm$	1.422	&	323.413	$\pm$	2.575	\\
55050.304115	&	61.227	$\pm$	4.059	&	79.092	$\pm$	0.294	&	15.909	$\pm$	1.874	&	32.386	$\pm$	0.386	&	64.700	$\pm$	1.411	&	320.649	$\pm$	2.736	\\
55053.767589	&	62.461	$\pm$	2.430	&	78.973	$\pm$	0.210	&	16.465	$\pm$	1.245	&	32.507	$\pm$	0.276	&	66.362	$\pm$	0.981	&	321.805	$\pm$	1.907	\\
55057.874710	&	59.698	$\pm$	3.869	&	79.232	$\pm$	0.199	&	15.005	$\pm$	1.542	&	33.013	$\pm$	0.265	&	66.069	$\pm$	1.045	&	324.048	$\pm$	1.938	\\
55061.634456	&	63.015	$\pm$	2.974	&	78.326	$\pm$	0.269	&	17.080	$\pm$	1.633	&	32.120	$\pm$	0.393	&	62.205	$\pm$	1.452	&	321.125	$\pm$	2.605	\\
55065.230727	&	35.995	$\pm$	2.974	&	77.523	$\pm$	0.798	&	9.311	$\pm$	1.633	&	30.464	$\pm$	0.826	&	61.544	$\pm$	3.690	&	328.098	$\pm$	4.977	\\
55067.182111	&	75.691	$\pm$	1.294	&	69.690	$\pm$	5.531	&	31.217	$\pm$	1.982	&	19.782	$\pm$	5.433	&	24.124	$\pm$	9.473	&	279.457	$\pm$	2.929	\\
55070.277756	&	63.922	$\pm$	3.508	&	78.844	$\pm$	0.169	&	15.607	$\pm$	1.909	&	33.771	$\pm$	0.352	&	56.672	$\pm$	1.575	&	325.965	$\pm$	2.571	\\
55074.425711	&	66.321	$\pm$	3.445	&	78.750	$\pm$	0.187	&	16.152	$\pm$	2.283	&	33.518	$\pm$	0.458	&	53.919	$\pm$	2.210	&	324.187	$\pm$	3.291	\\
55077.848279	&	67.671	$\pm$	4.066	&	77.458	$\pm$	0.357	&	17.269	$\pm$	3.311	&	32.052	$\pm$	0.865	&	50.804	$\pm$	1.324	&	321.033	$\pm$	5.382	\\
55083.824984	&	68.052	$\pm$	2.600	&	77.793	$\pm$	0.203	&	17.142	$\pm$	2.122	&	33.066	$\pm$	0.505	&	53.435	$\pm$	2.540	&	322.217	$\pm$	3.232	\\
55084.734264	&	66.361	$\pm$	5.317	&	77.680	$\pm$	0.478	&	17.179	$\pm$	3.827	&	32.598	$\pm$	0.952	&	52.764	$\pm$	1.324	&	318.453	$\pm$	2.377	\\
55087.278192	&	71.806	$\pm$	3.214	&	76.954	$\pm$	0.719	&	21.685	$\pm$	4.859	&	31.254	$\pm$	2.171	&	46.173	$\pm$	1.324	&	313.208	$\pm$	2.366	\\
55094.102831	&	61.631	$\pm$	6.295	&	75.903	$\pm$	0.504	&	15.308	$\pm$	2.932	&	31.042	$\pm$	0.669	&	55.341	$\pm$	2.659	&	320.091	$\pm$	3.583	\\
55096.421985	&	63.947	$\pm$	4.856	&	76.563	$\pm$	0.390	&	15.843	$\pm$	2.604	&	32.215	$\pm$	0.557	&	57.951	$\pm$	2.617	&	318.779	$\pm$	3.213	\\
55100.978555	&	57.876	$\pm$	5.887	&	77.649	$\pm$	0.170	&	12.960	$\pm$	1.872	&	34.039	$\pm$	0.236	&	62.181	$\pm$	1.227	&	324.958	$\pm$	1.532	\\
\hline
\end{tabular}
\end{center}
\end{table*}

\setcounter{table}{1}
\begin{table*}
\setlength{\tabcolsep}{0.2 pt}
\begin{center}
\caption{Continued.}
\begin{tabular}{ccccccc}
\hline
\hline
JD	&	$b_{1}$ (Latitude)	&	$b_{2}$ (Latitude)	&	$g_{1}$ (Radius)	&	$g_{2}$  (Radius)	&	$l_{1}$  (Longitude)	&	$l_{2}$   (Longitude)	\\
\hline		 	 	 						 	 	 						 	 	 					
55105.453390	&	58.804	$\pm$	5.542	&	76.923	$\pm$	0.338	&	15.505	$\pm$	2.264	&	32.774	$\pm$	0.450	&	53.787	$\pm$	1.688	&	314.621	$\pm$	2.377	\\
55108.089248	&	73.020	$\pm$	2.975	&	74.902	$\pm$	0.241	&	27.145	$\pm$	5.195	&	27.606	$\pm$	4.971	&	32.067	$\pm$	1.324	&	291.603	$\pm$	2.366	\\
55111.430042	&	58.736	$\pm$	3.466	&	75.829	$\pm$	0.241	&	15.596	$\pm$	1.416	&	31.835	$\pm$	0.305	&	51.917	$\pm$	1.171	&	313.049	$\pm$	1.653	\\
55116.763048	&	63.599	$\pm$	2.688	&	76.973	$\pm$	0.264	&	17.157	$\pm$	1.592	&	32.509	$\pm$	0.396	&	52.047	$\pm$	1.535	&	312.718	$\pm$	2.365	\\
55120.236651	&	67.100	$\pm$	2.534	&	77.592	$\pm$	0.382	&	19.611	$\pm$	2.222	&	32.561	$\pm$	0.719	&	47.943	$\pm$	2.692	&	308.686	$\pm$	4.336	\\
55122.484275	&	65.708	$\pm$	2.534	&	78.011	$\pm$	0.594	&	17.748	$\pm$	4.261	&	33.933	$\pm$	1.028	&	54.446	$\pm$	4.318	&	314.813	$\pm$	2.782	\\
55125.610515	&	68.529	$\pm$	2.427	&	78.667	$\pm$	0.360	&	20.282	$\pm$	2.677	&	33.143	$\pm$	0.921	&	47.553	$\pm$	3.319	&	312.377	$\pm$	2.022	\\
55128.225931	&	67.447	$\pm$	3.104	&	78.347	$\pm$	0.393	&	19.391	$\pm$	2.933	&	32.574	$\pm$	0.945	&	48.032	$\pm$	3.430	&	313.894	$\pm$	2.197	\\
55130.514420	&	61.196	$\pm$	6.095	&	75.843	$\pm$	0.615	&	16.127	$\pm$	2.980	&	30.903	$\pm$	0.782	&	59.660	$\pm$	2.665	&	322.497	$\pm$	4.008	\\
55132.710959	&	67.228	$\pm$	2.467	&	78.319	$\pm$	0.351	&	18.628	$\pm$	2.028	&	32.079	$\pm$	0.614	&	49.971	$\pm$	2.332	&	309.434	$\pm$	3.950	\\
55140.403958	&	68.362	$\pm$	1.572	&	78.515	$\pm$	0.253	&	19.713	$\pm$	1.621	&	31.138	$\pm$	0.598	&	42.595	$\pm$	2.093	&	307.516	$\pm$	3.901	\\
55150.589779	&	68.603	$\pm$	2.413	&	79.115	$\pm$	0.203	&	18.488	$\pm$	2.404	&	31.868	$\pm$	0.727	&	42.700	$\pm$	2.881	&	314.182	$\pm$	2.219	\\
55158.415607	&	67.166	$\pm$	3.511	&	77.979	$\pm$	0.500	&	19.066	$\pm$	3.177	&	30.642	$\pm$	1.132	&	41.285	$\pm$	3.827	&	308.543	$\pm$	2.145	\\
55161.858566	&	73.557	$\pm$	1.875	&	74.930	$\pm$	0.500	&	28.440	$\pm$	2.855	&	24.497	$\pm$	0.289	&	22.522	$\pm$	2.332	&	273.844	$\pm$	2.366	\\
55165.352611	&	70.567	$\pm$	3.037	&	78.492	$\pm$	0.562	&	22.987	$\pm$	4.487	&	30.529	$\pm$	0.325	&	32.892	$\pm$	2.093	&	295.508	$\pm$	1.653	\\
55169.888745	&	68.733	$\pm$	3.314	&	78.484	$\pm$	0.562	&	20.493	$\pm$	3.744	&	31.984	$\pm$	1.419	&	36.448	$\pm$	1.324	&	301.355	$\pm$	2.571	\\
55173.862975	&	68.021	$\pm$	4.607	&	78.484	$\pm$	0.715	&	19.730	$\pm$	4.602	&	32.198	$\pm$	1.587	&	35.569	$\pm$	1.324	&	299.308	$\pm$	3.291	\\
55176.887073	&	65.779	$\pm$	3.268	&	77.793	$\pm$	0.463	&	18.602	$\pm$	2.495	&	32.093	$\pm$	0.767	&	38.022	$\pm$	2.685	&	298.566	$\pm$	4.758	\\
55180.493518	&	63.531	$\pm$	4.174	&	77.176	$\pm$	0.476	&	17.253	$\pm$	2.451	&	31.224	$\pm$	0.674	&	37.329	$\pm$	2.391	&	297.495	$\pm$	3.965	\\
55187.440787	&	68.403	$\pm$	3.204	&	78.177	$\pm$	0.544	&	19.051	$\pm$	2.973	&	32.592	$\pm$	0.932	&	37.112	$\pm$	1.147	&	295.397	$\pm$	2.693	\\
55191.159632	&	69.524	$\pm$	3.215	&	78.457	$\pm$	0.610	&	20.404	$\pm$	3.621	&	32.022	$\pm$	1.342	&	29.811	$\pm$	1.449	&	290.715	$\pm$	2.285	\\
55194.786534	&	65.594	$\pm$	4.560	&	78.938	$\pm$	0.273	&	15.563	$\pm$	2.677	&	33.146	$\pm$	0.509	&	37.822	$\pm$	2.447	&	302.929	$\pm$	3.696	\\
55198.556476	&	67.491	$\pm$	3.688	&	79.212	$\pm$	0.268	&	16.620	$\pm$	2.751	&	32.992	$\pm$	0.616	&	36.101	$\pm$	2.754	&	303.617	$\pm$	2.496	\\
55203.368529	&	66.201	$\pm$	3.404	&	79.536	$\pm$	0.149	&	15.008	$\pm$	1.965	&	33.587	$\pm$	0.326	&	38.443	$\pm$	1.768	&	305.854	$\pm$	2.577	\\
55278.564869	&	69.191	$\pm$	2.174	&	77.595	$\pm$	0.736	&	21.329	$\pm$	2.214	&	30.962	$\pm$	0.977	&	31.706	$\pm$	3.075	&	279.055	$\pm$	2.056	\\
55282.110192	&	61.455	$\pm$	3.692	&	78.379	$\pm$	0.358	&	16.646	$\pm$	1.801	&	32.119	$\pm$	0.421	&	38.373	$\pm$	1.415	&	289.975	$\pm$	2.653	\\
55286.115289	&	68.067	$\pm$	2.204	&	78.100	$\pm$	0.702	&	21.246	$\pm$	2.111	&	31.863	$\pm$	0.875	&	33.457	$\pm$	2.514	&	278.883	$\pm$	2.760	\\
55290.477987	&	67.725	$\pm$	2.826	&	77.905	$\pm$	0.809	&	21.153	$\pm$	2.580	&	32.151	$\pm$	1.016	&	31.765	$\pm$	3.020	&	277.276	$\pm$	2.495	\\
55294.299182	&	66.081	$\pm$	2.682	&	77.905	$\pm$	0.629	&	20.212	$\pm$	2.098	&	32.634	$\pm$	0.728	&	29.837	$\pm$	2.170	&	274.620	$\pm$	4.009	\\
55297.916038	&	64.709	$\pm$	3.003	&	78.589	$\pm$	0.350	&	17.892	$\pm$	1.877	&	33.916	$\pm$	0.445	&	34.050	$\pm$	1.583	&	283.230	$\pm$	2.813	\\
55301.829192	&	59.677	$\pm$	3.514	&	78.421	$\pm$	0.271	&	16.143	$\pm$	1.513	&	33.575	$\pm$	0.300	&	33.796	$\pm$	1.050	&	282.969	$\pm$	1.885	\\
55305.967125	&	63.953	$\pm$	2.074	&	78.028	$\pm$	0.303	&	18.464	$\pm$	1.283	&	33.333	$\pm$	0.344	&	30.741	$\pm$	1.147	&	277.405	$\pm$	1.998	\\
55311.729577	&	64.647	$\pm$	2.103	&	77.422	$\pm$	0.488	&	20.267	$\pm$	1.502	&	32.553	$\pm$	0.528	&	29.781	$\pm$	1.449	&	272.577	$\pm$	2.740	\\
55314.723194	&	65.233	$\pm$	1.989	&	77.395	$\pm$	0.588	&	21.063	$\pm$	1.526	&	31.947	$\pm$	0.610	&	28.626	$\pm$	1.609	&	268.837	$\pm$	3.056	\\
55318.248101	&	64.094	$\pm$	2.395	&	78.449	$\pm$	0.471	&	19.721	$\pm$	1.567	&	32.508	$\pm$	0.499	&	29.410	$\pm$	1.307	&	270.048	$\pm$	2.801	\\
55321.507363	&	67.747	$\pm$	2.158	&	77.049	$\pm$	0.830	&	22.326	$\pm$	1.986	&	31.277	$\pm$	0.930	&	23.642	$\pm$	2.585	&	263.668	$\pm$	2.343	\\
55324.327288	&	63.231	$\pm$	1.828	&	77.392	$\pm$	0.361	&	19.259	$\pm$	1.119	&	32.241	$\pm$	0.360	&	30.961	$\pm$	1.027	&	270.101	$\pm$	1.825	\\
55327.893061	&	66.778	$\pm$	2.175	&	78.474	$\pm$	0.570	&	20.473	$\pm$	1.753	&	32.267	$\pm$	0.632	&	28.538	$\pm$	1.718	&	269.419	$\pm$	3.508	\\
\hline
\end{tabular}
\end{center}
\end{table*}

\setcounter{table}{1}
\begin{table*}
\setlength{\tabcolsep}{0.2 pt}
\begin{center}
\caption{Continued.}
\begin{tabular}{ccccccc}
\hline
\hline
JD	&	$b_{1}$ (Latitude)	&	$b_{2}$ (Latitude)	&	$g_{1}$ (Radius)	&	$g_{2}$  (Radius)	&	$l_{1}$  (Longitude)	&	$l_{2}$   (Longitude)	\\
\hline		 	 	 						 	 	 						 	 	 					
55331.346445	&	65.131	$\pm$	1.995	&	77.933	$\pm$	0.491	&	19.913	$\pm$	1.409	&	31.583	$\pm$	0.504	&	27.701	$\pm$	1.431	&	267.882	$\pm$	2.688	\\
55334.850912	&	61.023	$\pm$	3.036	&	77.925	$\pm$	0.411	&	17.550	$\pm$	1.484	&	31.577	$\pm$	0.394	&	27.811	$\pm$	1.149	&	269.490	$\pm$	2.184	\\
55339.489476	&	66.995	$\pm$	2.168	&	76.507	$\pm$	0.859	&	21.517	$\pm$	1.808	&	30.605	$\pm$	0.845	&	26.610	$\pm$	2.198	&	263.648	$\pm$	3.641	\\
55343.126760	&	66.924	$\pm$	1.603	&	76.246	$\pm$	0.673	&	22.089	$\pm$	1.338	&	30.500	$\pm$	0.656	&	24.992	$\pm$	1.652	&	260.579	$\pm$	2.749	\\
55347.601842	&	66.501	$\pm$	1.506	&	75.809	$\pm$	0.605	&	22.068	$\pm$	1.207	&	30.153	$\pm$	0.593	&	23.722	$\pm$	1.523	&	260.257	$\pm$	2.451	\\
55351.157383	&	61.440	$\pm$	2.749	&	77.325	$\pm$	0.496	&	18.424	$\pm$	1.453	&	31.620	$\pm$	0.446	&	28.428	$\pm$	1.200	&	266.097	$\pm$	2.191	\\
55353.486873	&	68.538	$\pm$	1.992	&	77.246	$\pm$	0.897	&	22.360	$\pm$	1.850	&	30.903	$\pm$	0.913	&	25.707	$\pm$	2.319	&	258.643	$\pm$	3.919	\\
55356.490687	&	68.026	$\pm$	2.091	&	78.397	$\pm$	0.784	&	21.722	$\pm$	1.892	&	31.396	$\pm$	0.835	&	24.741	$\pm$	2.067	&	260.950	$\pm$	4.201	\\
55359.494498	&	69.289	$\pm$	2.484	&	79.019	$\pm$	0.361	&	22.689	$\pm$	2.538	&	31.424	$\pm$	0.289	&	25.281	$\pm$	2.819	&	258.597	$\pm$	2.118	\\
55362.978506	&	70.732	$\pm$	1.760	&	78.142	$\pm$	0.570	&	24.305	$\pm$	1.956	&	30.223	$\pm$	0.325	&	23.033	$\pm$	2.644	&	255.656	$\pm$	2.010	\\
55366.636200	&	67.624	$\pm$	1.958	&	78.918	$\pm$	0.847	&	22.111	$\pm$	1.787	&	30.882	$\pm$	0.858	&	25.284	$\pm$	1.788	&	259.436	$\pm$	4.395	\\
55369.783038	&	65.885	$\pm$	2.325	&	79.667	$\pm$	0.717	&	20.771	$\pm$	1.808	&	31.341	$\pm$	0.721	&	25.963	$\pm$	1.545	&	261.395	$\pm$	4.223	\\
55373.723379	&	69.688	$\pm$	0.734	&	78.148	$\pm$	0.519	&	24.727	$\pm$	0.741	&	29.172	$\pm$	0.517	&	20.718	$\pm$	0.884	&	248.008	$\pm$	1.984	\\
55376.393757	&	67.823	$\pm$	0.835	&	78.289	$\pm$	0.447	&	22.743	$\pm$	0.775	&	28.870	$\pm$	0.459	&	20.538	$\pm$	0.897	&	256.363	$\pm$	2.215	\\
55379.845729	&	66.688	$\pm$	0.657	&	79.422	$\pm$	0.266	&	21.339	$\pm$	0.554	&	28.900	$\pm$	0.282	&	19.259	$\pm$	0.559	&	257.882	$\pm$	1.646	\\
55383.285777	&	66.523	$\pm$	0.764	&	79.697	$\pm$	0.297	&	21.045	$\pm$	0.628	&	28.839	$\pm$	0.311	&	19.361	$\pm$	0.608	&	257.566	$\pm$	1.866	\\
55387.551685	&	66.708	$\pm$	0.595	&	80.077	$\pm$	0.213	&	20.520	$\pm$	0.486	&	28.623	$\pm$	0.231	&	18.404	$\pm$	0.466	&	258.571	$\pm$	1.482	\\
55392.343077	&	66.536	$\pm$	0.711	&	80.006	$\pm$	0.264	&	20.452	$\pm$	0.566	&	27.872	$\pm$	0.283	&	14.864	$\pm$	0.561	&	254.923	$\pm$	1.804	\\
55396.930800	&	71.601	$\pm$	0.462	&	76.197	$\pm$	0.556	&	25.365	$\pm$	0.441	&	23.531	$\pm$	0.519	&	6.999	$\pm$	0.799	&	234.162	$\pm$	1.513	\\
55401.464365	&	74.165	$\pm$	0.778	&	75.475	$\pm$	1.243	&	26.908	$\pm$	0.738	&	21.316	$\pm$	1.208	&	3.240	$\pm$	1.728	&	230.587	$\pm$	2.751	\\
55404.659853	&	69.298	$\pm$	0.678	&	80.712	$\pm$	0.441	&	22.353	$\pm$	0.661	&	27.070	$\pm$	0.433	&	12.554	$\pm$	0.654	&	241.520	$\pm$	2.359	\\
55408.234381    &	71.552	$\pm$	0.702	&	79.246	$\pm$	0.707	&	24.458	$\pm$	0.692	&	25.333	$\pm$	0.641	&	10.011	$\pm$	0.757	&	226.661	$\pm$	2.048	\\
55410.950691	&	70.428	$\pm$	0.765	&	78.487	$\pm$	0.739	&	24.069	$\pm$	0.675	&	24.809	$\pm$	0.628	&	9.584	$\pm$	0.743	&	225.428	$\pm$	1.920	\\
55414.314060	&	67.766	$\pm$	0.758	&	82.063	$\pm$	0.363	&	20.547	$\pm$	0.635	&	27.598	$\pm$	0.327	&	13.545	$\pm$	0.460	&	238.909	$\pm$	2.124	\\
55417.755753	&	67.843	$\pm$	0.951	&	82.471	$\pm$	0.418	&	20.333	$\pm$	0.817	&	27.539	$\pm$	0.413	&	11.979	$\pm$	0.619	&	243.091	$\pm$	3.137	\\
55421.889452	&	68.636	$\pm$	0.774	&	83.132	$\pm$	0.284	&	19.923	$\pm$	0.717	&	27.831	$\pm$	0.338	&	14.122	$\pm$	0.559	&	253.879	$\pm$	3.132	\\
55426.614352	&	67.900	$\pm$	0.822	&	83.328	$\pm$	0.236	&	19.255	$\pm$	0.702	&	28.520	$\pm$	0.289	&	14.666	$\pm$	0.522	&	256.507	$\pm$	2.886	\\
55429.912649	&	69.051	$\pm$	1.108	&	83.615	$\pm$	0.400	&	20.039	$\pm$	1.077	&	29.060	$\pm$	0.470	&	15.990	$\pm$	0.770	&	253.600	$\pm$	4.560	\\
55433.240206	&	72.787	$\pm$	2.094	&	84.692	$\pm$	1.007	&	21.453	$\pm$	2.940	&	27.567	$\pm$	1.691	&	10.110	$\pm$	2.408	&	255.676	$\pm$	18.000	\\
55435.678588	&	72.312	$\pm$	0.851	&	81.306	$\pm$	0.802	&	24.269	$\pm$	0.965	&	26.261	$\pm$	0.811	&	11.136	$\pm$	0.960	&	231.213	$\pm$	3.429	\\
55438.429596	&	75.418	$\pm$	0.567	&	75.819	$\pm$	1.028	&	27.761	$\pm$	0.512	&	21.045	$\pm$	0.940	&	6.928	$\pm$	0.997	&	220.547	$\pm$	1.531	\\
55441.877717	&	72.414	$\pm$	0.732	&	81.364	$\pm$	0.679	&	23.981	$\pm$	0.821	&	26.171	$\pm$	0.677	&	11.250	$\pm$	0.800	&	230.361	$\pm$	2.831	\\
55445.513138	&	65.529	$\pm$	1.033	&	84.504	$\pm$	0.178	&	17.364	$\pm$	0.663	&	29.201	$\pm$	0.202	&	14.208	$\pm$	0.373	&	254.044	$\pm$	2.528	\\
55448.922765	&	67.560	$\pm$	0.842	&	83.195	$\pm$	0.306	&	19.168	$\pm$	0.660	&	27.787	$\pm$	0.284	&	12.294	$\pm$	0.436	&	240.545	$\pm$	2.369	\\
55452.903836	&	71.702	$\pm$	0.642	&	79.618	$\pm$	0.654	&	23.844	$\pm$	0.611	&	24.590	$\pm$	0.570	&	9.030	$\pm$	0.615	&	221.961	$\pm$	1.739	\\
55458.651004	&	67.454	$\pm$	0.602	&	83.186	$\pm$	0.298	&	20.039	$\pm$	0.475	&	27.345	$\pm$	0.234	&	13.438	$\pm$	0.248	&	225.998	$\pm$	1.374	\\
55464.327988	&	64.347	$\pm$	1.018	&	83.193	$\pm$	0.313	&	18.492	$\pm$	0.611	&	27.115	$\pm$	0.251	&	13.667	$\pm$	0.346	&	230.084	$\pm$	1.759	\\
55467.113359	&	75.280	$\pm$	0.418	&	71.792	$\pm$	0.986	&	27.533	$\pm$	0.285	&	18.648	$\pm$	0.678	&	5.116	$\pm$	0.399	&	203.328	$\pm$	0.491	\\
55470.866573	&	71.397	$\pm$	0.516	&	78.158	$\pm$	0.634	&	24.780	$\pm$	0.436	&	23.353	$\pm$	0.497	&	9.349	$\pm$	0.344	&	210.027	$\pm$	0.862	\\
55474.258800	&	67.772	$\pm$	0.884	&	82.891	$\pm$	0.587	&	21.131	$\pm$	0.732	&	26.546	$\pm$	0.438	&	11.610	$\pm$	0.334	&	215.610	$\pm$	1.840	\\
\hline
\end{tabular}
\end{center}
\end{table*}

\setcounter{table}{1}
\begin{table*}
\setlength{\tabcolsep}{0.2 pt}
\begin{center}
\caption{Continued.}
\begin{tabular}{ccccccc}
\hline
\hline
JD	&	$b_{1}$ (Latitude)	&	$b_{2}$ (Latitude)	&	$g_{1}$ (Radius)	&	$g_{2}$  (Radius)	&	$l_{1}$  (Longitude)	&	$l_{2}$   (Longitude)	\\
\hline		 	 	 						 	 	 						 	 	 					
55477.719474	&	67.483	$\pm$	0.709	&	81.740	$\pm$	0.445	&	20.901	$\pm$	0.544	&	26.305	$\pm$	0.323	&	14.101	$\pm$	0.265	&	215.318	$\pm$	1.092	\\
55482.134025	&	70.552	$\pm$	0.486	&	76.821	$\pm$	0.567	&	23.782	$\pm$	0.370	&	22.756	$\pm$	0.410	&	10.330	$\pm$	0.302	&	209.804	$\pm$	0.649	\\
55486.917386	&	71.526	$\pm$	0.541	&	78.525	$\pm$	0.583	&	23.993	$\pm$	0.459	&	23.859	$\pm$	0.461	&	8.352	$\pm$	0.355	&	210.133	$\pm$	0.868	\\
55491.373477	&	73.059	$\pm$	0.638	&	75.799	$\pm$	0.844	&	24.945	$\pm$	0.517	&	22.085	$\pm$	0.669	&	8.792	$\pm$	0.687	&	214.372	$\pm$	1.077	\\
55495.871451	&	73.925	$\pm$	0.403	&	70.787	$\pm$	0.776	&	25.801	$\pm$	0.260	&	19.198	$\pm$	0.499	&	7.578	$\pm$	0.319	&	204.576	$\pm$	0.374	\\
55499.097473	&	72.097	$\pm$	0.543	&	74.413	$\pm$	0.689	&	24.582	$\pm$	0.388	&	22.083	$\pm$	0.490	&	10.643	$\pm$	0.318	&	205.162	$\pm$	0.476	\\
55501.742858	&	73.291	$\pm$	0.506	&	71.403	$\pm$	0.825	&	25.372	$\pm$	0.339	&	20.259	$\pm$	0.550	&	7.233	$\pm$	0.426	&	204.146	$\pm$	0.470	\\
55504.467590	&	72.524	$\pm$	0.462	&	71.336	$\pm$	0.772	&	25.018	$\pm$	0.297	&	20.040	$\pm$	0.491	&	9.101	$\pm$	0.280	&	203.230	$\pm$	0.385	\\
55507.360894	&	67.462	$\pm$	0.627	&	76.505	$\pm$	0.487	&	21.467	$\pm$	0.401	&	23.736	$\pm$	0.317	&	13.054	$\pm$	0.234	&	208.283	$\pm$	0.462	\\
55510.117978	&	71.784	$\pm$	0.618	&	73.560	$\pm$	0.827	&	23.813	$\pm$	0.418	&	21.027	$\pm$	0.554	&	11.875	$\pm$	0.375	&	207.317	$\pm$	0.535	\\
55512.869273	&	70.697	$\pm$	0.476	&	72.657	$\pm$	0.551	&	23.350	$\pm$	0.307	&	21.827	$\pm$	0.355	&	11.878	$\pm$	0.250	&	205.546	$\pm$	0.331	\\
55516.326187	&	69.021	$\pm$	0.592	&	77.486	$\pm$	0.453	&	21.440	$\pm$	0.416	&	24.709	$\pm$	0.303	&	13.343	$\pm$	0.212	&	205.681	$\pm$	0.400	\\
55519.665612	&	69.159	$\pm$	0.778	&	77.485	$\pm$	0.562	&	21.107	$\pm$	0.540	&	24.592	$\pm$	0.380	&	13.768	$\pm$	0.289	&	205.216	$\pm$	0.448	\\
55522.058312	&	71.307	$\pm$	0.671	&	73.982	$\pm$	0.728	&	22.541	$\pm$	0.450	&	22.173	$\pm$	0.476	&	12.652	$\pm$	0.295	&	202.894	$\pm$	0.441	\\
55524.814719	&	71.323	$\pm$	0.815	&	73.845	$\pm$	0.836	&	21.754	$\pm$	0.529	&	21.647	$\pm$	0.535	&	12.264	$\pm$	0.379	&	203.441	$\pm$	0.472	\\
55527.451254	&	72.219	$\pm$	0.551	&	74.986	$\pm$	0.485	&	21.968	$\pm$	0.381	&	23.235	$\pm$	0.334	&	15.422	$\pm$	0.205	&	201.570	$\pm$	0.232	\\
55530.126272	&	78.627	$\pm$	0.245	&	-1.110	$\pm$	98.070	&	26.256	$\pm$	0.134	&	8.622	$\pm$	0.427	&	10.499	$\pm$	0.479	&	186.390	$\pm$	0.446	\\
55532.172637	&	73.401	$\pm$	0.513	&	71.201	$\pm$	0.652	&	22.975	$\pm$	0.322	&	20.574	$\pm$	0.411	&	15.939	$\pm$	0.219	&	199.223	$\pm$	0.223	\\
55534.780911	&	71.429	$\pm$	0.603	&	73.233	$\pm$	0.630	&	22.334	$\pm$	0.388	&	21.695	$\pm$	0.405	&	17.332	$\pm$	0.239	&	198.730	$\pm$	0.245	\\
55536.878020	&	69.268	$\pm$	0.866	&	74.579	$\pm$	0.801	&	21.263	$\pm$	0.535	&	22.150	$\pm$	0.498	&	20.782	$\pm$	0.238	&	202.037	$\pm$	0.371	\\
55538.400281	&	72.296	$\pm$	0.638	&	70.433	$\pm$	1.057	&	23.209	$\pm$	0.381	&	18.925	$\pm$	0.606	&	18.506	$\pm$	0.324	&	207.543	$\pm$	0.393	\\
55541.482607	&	72.195	$\pm$	0.414	&	73.048	$\pm$	0.475	&	22.644	$\pm$	0.270	&	21.419	$\pm$	0.307	&	15.201	$\pm$	0.162	&	199.068	$\pm$	0.194	\\
55546.558864	&	72.005	$\pm$	0.557	&	74.864	$\pm$	0.465	&	22.111	$\pm$	0.394	&	24.176	$\pm$	0.323	&	16.019	$\pm$	0.208	&	202.905	$\pm$	0.235	\\
55550.658428	&	73.082	$\pm$	0.560	&	74.177	$\pm$	0.524	&	22.881	$\pm$	0.399	&	23.505	$\pm$	0.373	&	14.069	$\pm$	0.245	&	200.617	$\pm$	0.240	\\
56421.907557	&	74.336	$\pm$	0.319	&	112.006	$\pm$	0.807	&	26.349	$\pm$	0.199	&	18.653	$\pm$	0.451	&	102.414	$\pm$	0.212	&	111.379	$\pm$	0.257	\\
\hline
\end{tabular}
\end{center}
\end{table*}

\setcounter{table}{2}
\begin{table*}
\begin{center}
\caption{The parameters calculated for each flare detected with analysis of the Short Cadence data obtained by Kepler Mission for KOI-256 are listed.}
\begin{tabular}{cccccc}
\hline
\hline
$JD$	&	$P$	&	$T_{r}$	&	$T_{d}$	&	$T_{t}$	&	$Amplitude$	\\
(+24 00000)	&	(s)	&	(s)	&	(s)	&	(s)	&	(Intensity)	\\
\hline	 	 	 	 	 	 	 	 	 	 	
55375.342090	&	44.20740	&	235.39162	&	2353.99133	&	2589.38294	&	0.07436	\\
55375.947616	&	16.56043	&	176.54371	&	1118.15597	&	1294.69968	&	0.02621	\\
55376.350846	&	15.26603	&	411.94483	&	765.05904	&	1177.00387	&	0.04393	\\
55376.476174	&	22.31022	&	58.84790	&	1176.99523	&	1235.84314	&	0.05763	\\
55377.033340	&	12.30394	&	176.55322	&	1176.98573	&	1353.53894	&	0.02639	\\
55377.296257	&	27.67483	&	588.49718	&	882.75485	&	1471.25203	&	0.03511	\\
55378.175597	&	20.63149	&	176.55235	&	2236.29206	&	2412.84442	&	0.03071	\\
55378.328852	&	1.99895	&	58.85654	&	235.39248	&	294.24902	&	0.01837	\\
55380.097067	&	34.60774	&	235.40026	&	2824.77888	&	3060.17914	&	0.02929	\\
55380.590205	&	7.63004	&	58.84790	&	765.05731	&	823.90522	&	0.02825	\\
55380.741416	&	8.25863	&	294.24816	&	823.90522	&	1118.15338	&	0.01060	\\
55381.267930	&	19.27064	&	235.40026	&	1176.99264	&	1412.39290	&	0.04652	\\
55381.637784	&	7.80086	&	529.63978	&	647.34336	&	1176.98314	&	0.01822	\\
55383.117879	&	1.83830	&	58.83926	&	58.84704	&	117.68630	&	0.03887	\\
55383.318131	&	1.72242	&	58.84790	&	58.85654	&	117.70445	&	0.02840	\\
55385.586972	&	6.76596	&	1176.99005	&	235.39939	&	1412.38944	&	0.00939	\\
55385.605363	&	15.97846	&	1176.99005	&	2118.58157	&	3295.57162	&	0.01176	\\
55386.161845	&	1.38046	&	176.54285	&	117.70445	&	294.24730	&	0.01294	\\
55386.294665	&	9.55332	&	176.56099	&	1353.53290	&	1530.09389	&	0.02204	\\
55386.812322	&	10.15885	&	1059.29424	&	823.90349	&	1883.19773	&	0.01717	\\
55389.135650	&	15.95644	&	1412.38771	&	1176.98832	&	2589.37603	&	0.01390	\\
55390.912711	&	8.01176	&	470.79878	&	823.90176	&	1294.70054	&	0.02250	\\
55391.021692	&	4.32847	&	353.08570	&	353.09520	&	706.18090	&	0.01673	\\
55391.457613	&	9.42058	&	117.70358	&	706.19731	&	823.90090	&	0.02878	\\
55392.580110	&	4.03619	&	117.70358	&	470.79792	&	588.50150	&	0.01606	\\
55392.744261	&	9.80095	&	294.24643	&	1471.23302	&	1765.47946	&	0.01100	\\
55393.700563	&	23.57782	&	235.39853	&	2118.56429	&	2353.96282	&	0.03602	\\
55394.247507	&	4.93193	&	117.70358	&	529.64496	&	647.34854	&	0.02670	\\
55395.160216	&	13.90038	&	117.70358	&	706.19558	&	823.89917	&	0.04806	\\
55395.578427	&	404.82960	&	1118.12832	&	13417.62970	&	14535.75802	&	0.27309	\\
55397.006747	&	38.35936	&	765.04262	&	2118.56947	&	2883.61210	&	0.03148	\\
55401.269904	&	42.89049	&	117.69408	&	1412.37994	&	1530.07402	&	0.13944	\\
55404.761341	&	7.68292	&	176.54198	&	765.04003	&	941.58202	&	0.03032	\\
55405.599122	&	3.58347	&	117.70272	&	353.09174	&	470.79446	&	0.01701	\\
55406.776100	&	7.62807	&	176.54976	&	529.63373	&	706.18349	&	0.03081	\\
55407.746018	&	9.85367	&	176.54198	&	823.88534	&	1000.42733	&	0.02354	\\
55408.136300	&	58.03656	&	470.79360	&	3060.13853	&	3530.93213	&	0.05706	\\
\hline
\end{tabular}
\end{center}
\end{table*}

\setcounter{table}{2}
\begin{table*}
\begin{center}
\caption{Continued.}
\begin{tabular}{cccccc}
\hline
\hline
$JD$	&	$P$	&	$T_{r}$	&	$T_{d}$	&	$T_{t}$	&	$Amplitude$	\\
(+24 00000)	&	(s)	&	(s)	&	(s)	&	(s)	&	(Intensity)	\\
\hline
55409.594580	&	75.45501	&	235.38902	&	2824.75555	&	3060.14458	&	0.13119	\\
55409.832291	&	34.40761	&	176.54976	&	3001.29667	&	3177.84643	&	0.04241	\\
55409.890187	&	26.17318	&	117.70272	&	1824.31181	&	1942.01453	&	0.05453	\\
55412.681417	&	50.01182	&	294.23434	&	2765.89642	&	3060.13075	&	0.05973	\\
55412.876899	&	6.91584	&	176.54976	&	588.47904	&	765.02880	&	0.02533	\\
55413.391825	&	6.37731	&	58.84704	&	529.63978	&	588.48682	&	0.02509	\\
55414.867810	&	5.48103	&	117.70186	&	588.49546	&	706.19731	&	0.02086	\\
55414.925024	&	12.11842	&	235.39680	&	1294.67549	&	1530.07229	&	0.01774	\\
55416.699338	&	12.14917	&	176.54112	&	1647.76464	&	1824.30576	&	0.02079	\\
55417.073953	&	4.53704	&	176.54112	&	647.33299	&	823.87411	&	0.01461	\\
55417.940336	&	2.68543	&	58.84704	&	294.24298	&	353.09002	&	0.01618	\\
55418.479781	&	7.96692	&	294.24298	&	1000.42214	&	1294.66512	&	0.01591	\\
55418.700463	&	40.81725	&	1353.52080	&	2589.33802	&	3942.85882	&	0.03181	\\
55418.793776	&	4.98763	&	176.54026	&	588.48595	&	765.02621	&	0.01339	\\
55419.427897	&	82.95145	&	823.87238	&	4237.09920	&	5060.97158	&	0.04302	\\
55419.497371	&	185.47791	&	1118.12400	&	8827.28755	&	9945.41155	&	0.07175	\\
55421.411309	&	8.11929	&	235.38730	&	941.58288	&	1176.97018	&	0.02561	\\
55423.293915	&	68.96291	&	353.08915	&	5237.51011	&	5590.59926	&	0.05211	\\
55425.451690	&	5.81840	&	58.85568	&	588.48336	&	647.33904	&	0.02426	\\
55425.581783	&	11.64399	&	941.58029	&	647.33040	&	1588.91069	&	0.02157	\\
55427.109523	&	10.32274	&	235.38643	&	1000.42646	&	1235.81290	&	0.02436	\\
55428.162527	&	6.31033	&	117.69322	&	470.78150	&	588.47472	&	0.02500	\\
55429.939553	&	14.85190	&	294.24125	&	1294.66598	&	1588.90723	&	0.03114	\\
55431.149892	&	6.49379	&	117.70186	&	882.71424	&	1000.41610	&	0.02000	\\
55431.203700	&	14.69386	&	411.95088	&	1294.65648	&	1706.60736	&	0.02178	\\
55433.127822	&	1.64001	&	176.53939	&	117.70099	&	294.24038	&	0.01724	\\
55433.193209	&	5.77048	&	58.84704	&	1176.96240	&	1235.80944	&	0.01492	\\
55434.151534	&	1.12874	&	176.55667	&	58.83754	&	235.39421	&	0.01448	\\
55434.157664	&	2.97310	&	117.69235	&	411.94224	&	529.63459	&	0.01664	\\
55434.353825	&	6.63105	&	117.68371	&	1588.90378	&	1706.58749	&	0.01544	\\
55435.784160	&	4.06328	&	117.70099	&	529.63459	&	647.33558	&	0.02182	\\
55435.835924	&	0.73911	&	58.84618	&	58.85482	&	117.70099	&	0.01707	\\
55435.926512	&	3.61430	&	235.40198	&	706.17312	&	941.57510	&	0.01044	\\
55436.045026	&	7.41432	&	58.84618	&	1294.65389	&	1353.50006	&	0.01989	\\
55436.068183	&	3.51315	&	117.68371	&	529.64237	&	647.32608	&	0.01205	\\
55436.081125	&	2.71089	&	353.08656	&	588.48077	&	941.56733	&	0.00666	\\
55436.558584	&	23.75906	&	353.08656	&	2236.22035	&	2589.30691	&	0.01960	\\
55436.636912	&	7.82704	&	353.08570	&	823.87411	&	1176.95981	&	0.00867	\\
55436.794249	&	11.70227	&	176.54717	&	882.71165	&	1059.25882	&	0.03533	\\
55438.701360	&	21.11367	&	117.69235	&	3413.18621	&	3530.87856	&	0.03614	\\
\hline
\end{tabular}
\end{center}
\end{table*}

\setcounter{table}{2}
\begin{table*}
\begin{center}
\caption{Continued.}
\begin{tabular}{cccccc}
\hline
\hline
$JD$	&	$P$	&	$T_{r}$	&	$T_{d}$	&	$T_{t}$	&	$Amplitude$	\\
(+24 00000)	&	(s)	&	(s)	&	(s)	&	(s)	&	(Intensity)	\\
\hline
55438.913866	&	5.61630	&	353.08570	&	1118.11363	&	1471.19933	&	0.01361	\\
55441.226917	&	32.58644	&	647.32522	&	2412.76147	&	3060.08669	&	0.04902	\\
55441.943445	&	3.33076	&	235.38470	&	58.85482	&	294.23952	&	0.02369	\\
55443.211671	&	6.27441	&	235.39335	&	941.57251	&	1176.96586	&	0.01589	\\
55443.229380	&	2.03953	&	470.78582	&	470.77805	&	941.56387	&	0.00682	\\
55443.527706	&	0.95940	&	58.84618	&	117.70099	&	176.54717	&	0.01647	\\
55443.531792	&	11.15569	&	235.39248	&	1294.64957	&	1530.04205	&	0.02500	\\
55445.415740	&	11.63756	&	235.39248	&	1059.26400	&	1294.65648	&	0.02883	\\
55446.473501	&	11.65998	&	941.56214	&	2000.82614	&	2942.38829	&	0.01389	\\
55446.954364	&	23.94391	&	529.63200	&	1883.12429	&	2412.75629	&	0.03194	\\
55447.452254	&	37.09812	&	1176.94598	&	3942.80266	&	5119.74864	&	0.01531	\\
55450.550613	&	26.19646	&	235.39248	&	2353.90579	&	2589.29827	&	0.04492	\\
55450.614637	&	15.81569	&	1118.10672	&	1294.65389	&	2412.76061	&	0.02139	\\
55451.372709	&	5.64617	&	117.70099	&	353.08397	&	470.78496	&	0.03361	\\
55453.086373	&	20.35148	&	176.55494	&	765.01325	&	941.56819	&	0.07455	\\
55453.672806	&	149.01479	&	235.38384	&	3118.92422	&	3354.30806	&	0.18088	\\
55456.080515	&	72.95417	&	529.63805	&	3942.77933	&	4472.41738	&	0.07085	\\
55456.193579	&	12.16574	&	176.54544	&	941.55869	&	1118.10413	&	0.03073	\\
55456.308004	&	15.39212	&	117.69149	&	1530.02477	&	1647.71626	&	0.02762	\\
55456.611096	&	5.30449	&	117.69149	&	470.77546	&	588.46694	&	0.02077	\\
55456.622675	&	12.84198	&	235.38298	&	2118.51763	&	2353.90061	&	0.01174	\\
55457.468608	&	193.53680	&	411.92842	&	8827.13117	&	9239.05958	&	0.11971	\\
55458.638065	&	6.67818	&	235.38298	&	882.72029	&	1118.10326	&	0.01673	\\
55460.915679	&	9.38679	&	58.84531	&	765.02016	&	823.86547	&	0.02893	\\
55460.929301	&	3.05990	&	117.70013	&	529.62854	&	647.32867	&	0.00933	\\
55461.137038	&	10.42624	&	58.84531	&	1000.41091	&	1059.25622	&	0.03052	\\
55461.401988	&	84.63194	&	176.54630	&	5002.01914	&	5178.56544	&	0.10579	\\
55462.177765	&	11.80962	&	353.08224	&	1118.11104	&	1471.19328	&	0.02530	\\
55463.505896	&	16.78388	&	117.70013	&	823.85597	&	941.55610	&	0.05735	\\
55464.891260	&	15.02389	&	1471.18291	&	1353.49142	&	2824.67434	&	0.01315	\\
55465.276084	&	10.70152	&	117.68198	&	1294.65475	&	1412.33674	&	0.02691	\\
55465.425926	&	14.17575	&	176.54544	&	1176.93734	&	1353.48278	&	0.04345	\\
55465.482458	&	18.31911	&	235.39939	&	1530.01958	&	1765.41898	&	0.05152	\\
55465.952419	&	41.43259	&	58.84618	&	58.84531	&	117.69149	&	0.68383	\\
55467.303046	&	68.23197	&	117.70013	&	3060.06336	&	3177.76349	&	0.13012	\\
55469.090940	&	10.93777	&	176.53680	&	941.55350	&	1118.09030	&	0.03609	\\
55469.613346	&	18.50133	&	823.85424	&	2530.43395	&	3354.28819	&	0.01302	\\
55470.019964	&	6.44893	&	117.70013	&	823.85338	&	941.55350	&	0.02676	\\
55470.607755	&	4.84475	&	117.69926	&	823.85424	&	941.55350	&	0.01444	\\
55471.355605	&	15.53370	&	235.39075	&	1176.93475	&	1412.32550	&	0.03245	\\
\hline
\end{tabular}
\end{center}
\end{table*}

\setcounter{table}{2}
\begin{table*}
\begin{center}
\caption{Continued.}
\begin{tabular}{cccccc}
\hline
\hline
$JD$	&	$P$	&	$T_{r}$	&	$T_{d}$	&	$T_{t}$	&	$Amplitude$	\\
(+24 00000)	&	(s)	&	(s)	&	(s)	&	(s)	&	(Intensity)	\\
\hline
55472.217880	&	19.35750	&	176.54458	&	3883.92019	&	4060.46477	&	0.01227	\\
55472.319364	&	5.77485	&	411.93533	&	882.69869	&	1294.63402	&	0.01891	\\
55472.813844	&	16.21783	&	117.70013	&	882.70733	&	1000.40746	&	0.05861	\\
55473.593024	&	51.78851	&	235.39075	&	5531.62435	&	5767.01510	&	0.03109	\\
55474.398085	&	12.41232	&	294.22742	&	1294.64179	&	1588.86922	&	0.02210	\\
55474.727057	&	44.88099	&	176.52730	&	1941.95923	&	2118.48653	&	0.18575	\\
55474.900057	&	9.91234	&	176.54458	&	941.55264	&	1118.09722	&	0.03113	\\
55475.423824	&	5.64696	&	117.69926	&	588.47126	&	706.17053	&	0.02536	\\
55475.562087	&	6.45265	&	58.83667	&	706.17053	&	765.00720	&	0.03652	\\
55476.091984	&	29.14921	&	235.38989	&	1353.48710	&	1588.87699	&	0.08591	\\
55476.294952	&	7.44200	&	58.84531	&	941.56042	&	1000.40573	&	0.02985	\\
55476.567392	&	314.03759	&	706.16189	&	21479.19898	&	22185.36086	&	0.07990	\\
55476.696120	&	21.90776	&	529.61731	&	1883.11219	&	2412.72950	&	0.02629	\\
55476.823486	&	86.57508	&	117.69926	&	8650.51920	&	8768.21846	&	0.03453	\\
55478.800721	&	21.97714	&	765.00634	&	1647.72144	&	2412.72778	&	0.02521	\\
55479.075204	&	48.89291	&	235.39853	&	2000.80195	&	2236.20048	&	0.08979	\\
55479.596246	&	75.63241	&	411.92582	&	4413.52886	&	4825.45469	&	0.06286	\\
55480.878077	&	3.63981	&	235.38125	&	706.16966	&	941.55091	&	0.00691	\\
55480.897829	&	53.97274	&	176.54458	&	2530.42618	&	2706.97075	&	0.07388	\\
55480.934608	&	46.92281	&	294.23520	&	4236.99120	&	4531.22640	&	0.02365	\\
55481.363701	&	147.57825	&	470.77978	&	5001.99754	&	5472.77731	&	0.07048	\\
55481.384815	&	160.06194	&	2295.02765	&	4295.83565	&	6590.86330	&	0.07118	\\
55482.448012	&	25.10970	&	235.38989	&	1588.86576	&	1824.25565	&	0.07721	\\
55484.817558	&	7.04579	&	117.69062	&	1000.40400	&	1118.09462	&	0.03365	\\
55486.206319	&	8.58246	&	176.52730	&	941.55869	&	1118.08598	&	0.02701	\\
55486.736215	&	5.60687	&	58.84531	&	588.47818	&	647.32349	&	0.01998	\\
55487.396200	&	39.15742	&	1941.95318	&	5119.68989	&	7061.64307	&	0.01433	\\
55488.308191	&	75.29180	&	353.07965	&	3530.81722	&	3883.89686	&	0.07725	\\
55488.373577	&	9.90490	&	529.62422	&	1883.10701	&	2412.73123	&	0.01054	\\
55489.807290	&	140.09412	&	3942.74909	&	9650.89814	&	13593.64723	&	0.02690	\\
55490.681140	&	1049.43937	&	2471.56790	&	15947.53056	&	18419.09846	&	0.33241	\\
55493.194394	&	14.87307	&	176.53594	&	647.32262	&	823.85856	&	0.04763	\\
55494.570213	&	54.05090	&	353.07965	&	2059.64986	&	2412.72950	&	0.11057	\\
55494.645815	&	28.36075	&	941.54832	&	2471.57482	&	3413.12314	&	0.01765	\\
55498.618661	&	9.03055	&	176.54371	&	1471.17168	&	1647.71539	&	0.01734	\\
55498.702436	&	3.62009	&	176.54371	&	470.77891	&	647.32262	&	0.01493	\\
55498.718101	&	13.40445	&	588.46867	&	2295.02074	&	2883.48941	&	0.01675	\\
55499.032768	&	5.29147	&	176.54458	&	823.84906	&	1000.39363	&	0.00952	\\
55500.394283	&	15.19193	&	117.68198	&	1706.55984	&	1824.24182	&	0.03291	\\
55501.954679	&	6.79482	&	58.85395	&	1118.09117	&	1176.94512	&	0.02542	\\
\hline
\end{tabular}
\end{center}
\end{table*}

\setcounter{table}{2}
\begin{table*}
\begin{center}
\caption{Continued.}
\begin{tabular}{cccccc}
\hline
\hline
$JD$	&	$P$	&	$T_{r}$	&	$T_{d}$	&	$T_{t}$	&	$Amplitude$	\\
(+24 00000)	&	(s)	&	(s)	&	(s)	&	(s)	&	(Intensity)	\\
\hline
55502.011891	&	18.73455	&	588.46781	&	2236.18406	&	2824.65187	&	0.01551	\\
55502.410333	&	11.46807	&	176.53507	&	765.01238	&	941.54746	&	0.03713	\\
55504.495855	&	9.53926	&	117.69840	&	1118.08339	&	1235.78179	&	0.02625	\\
55505.411932	&	55.53928	&	176.54371	&	2883.49718	&	3060.04090	&	0.08309	\\
55505.476636	&	60.13398	&	117.69840	&	5001.98198	&	5119.68038	&	0.10543	\\
55508.166292	&	1.79399	&	58.84531	&	58.85395	&	117.69926	&	0.03583	\\
55508.891661	&	14.28894	&	235.38902	&	1000.40141	&	1235.79043	&	0.04915	\\
55509.218588	&	21.62582	&	706.16794	&	2706.95347	&	3413.12141	&	0.01718	\\
55509.391587	&	31.00352	&	294.24298	&	2824.65274	&	3118.89571	&	0.03682	\\
55510.390758	&	36.78295	&	176.55235	&	4236.96960	&	4413.52195	&	0.05801	\\
55510.739480	&	11.95360	&	882.70214	&	941.54832	&	1824.25046	&	0.02207	\\
55511.376307	&	51.67114	&	588.46867	&	3177.72288	&	3766.19155	&	0.05542	\\
55513.722008	&	9.63831	&	176.53507	&	1883.10442	&	2059.63949	&	0.01544	\\
55514.407193	&	8.84904	&	117.69062	&	1471.17946	&	1588.87008	&	0.02198	\\
55515.838861	&	13.31654	&	235.38902	&	1941.94195	&	2177.33098	&	0.01942	\\
55516.291791	&	11.92080	&	353.07965	&	1353.48106	&	1706.56070	&	0.02661	\\
55516.392594	&	39.99328	&	3118.89658	&	2177.32234	&	5296.21891	&	0.01812	\\
55516.430735	&	23.22504	&	117.68976	&	2706.95434	&	2824.64410	&	0.02021	\\
55517.760920	&	98.70816	&	3177.73325	&	6826.23677	&	10003.97002	&	0.02622	\\
55520.294607	&	67.24819	&	176.54371	&	6708.53923	&	6885.08294	&	0.03740	\\
55522.737707	&	39.92723	&	176.53594	&	2236.18666	&	2412.72259	&	0.03204	\\
55522.816034	&	75.88575	&	176.53507	&	2177.34134	&	2353.87642	&	0.19217	\\
55524.700634	&	23.03246	&	58.84531	&	2353.87728	&	2412.72259	&	0.03093	\\
55525.479130	&	50.79022	&	353.08051	&	2765.81174	&	3118.89226	&	0.09610	\\
55526.711920	&	9.11512	&	117.69926	&	588.47818	&	706.17744	&	0.03393	\\
55527.096741	&	8.51295	&	117.69840	&	1235.77661	&	1353.47501	&	0.01772	\\
55527.229556	&	7.92308	&	353.07965	&	1176.93907	&	1530.01872	&	0.01985	\\
55527.828923	&	17.13460	&	176.53507	&	2059.64467	&	2236.17974	&	0.03384	\\
55528.409901	&	6.92449	&	176.54458	&	647.31485	&	823.85942	&	0.01744	\\
55530.246145	&	602.45858	&	353.08051	&	5060.84976	&	5413.93027	&	0.62493	\\
55530.609853	&	110.84606	&	529.62509	&	2412.72605	&	2942.35114	&	0.13490	\\
55530.638459	&	22.64046	&	58.84531	&	1000.40400	&	1059.24931	&	0.05504	\\
55531.639676	&	20.11803	&	823.86029	&	765.01498	&	1588.87526	&	0.03575	\\
55532.322138	&	5.99543	&	58.84531	&	941.55091	&	1000.39622	&	0.02514	\\
55532.377988	&	17.43303	&	117.69926	&	882.69696	&	1000.39622	&	0.04528	\\
55532.540771	&	19.95533	&	117.69926	&	1412.32205	&	1530.02131	&	0.05189	\\
55534.100491	&	11.44830	&	117.69926	&	1235.78698	&	1353.48624	&	0.01969	\\
55534.874221	&	7.51594	&	117.70013	&	765.00634	&	882.70646	&	0.03108	\\
55535.983734	&	69.75887	&	235.38125	&	3707.37907	&	3942.76032	&	0.07953	\\
55536.041628	&	40.06649	&	117.69926	&	2706.97334	&	2824.67261	&	0.04786	\\
\hline
\end{tabular}
\end{center}
\end{table*}

\setcounter{table}{2}
\begin{table*}
\begin{center}
\caption{Continued.}
\begin{tabular}{cccccc}
\hline
\hline
$JD$	&	$P$	&	$T_{r}$	&	$T_{d}$	&	$T_{t}$	&	$Amplitude$	\\
(+24 00000)	&	(s)	&	(s)	&	(s)	&	(s)	&	(Intensity)	\\
\hline
55536.117911	&	27.10988	&	117.69149	&	1235.78698	&	1353.47846	&	0.07168	\\
55536.827618	&	59.17967	&	882.70646	&	3354.28214	&	4236.98861	&	0.03548	\\
55538.078801	&	357.68647	&	176.54458	&	2589.26717	&	2765.81174	&	0.71509	\\
55538.144186	&	40.92859	&	117.69926	&	2059.64986	&	2177.34912	&	0.07322	\\
55539.366763	&	7.80770	&	117.69062	&	706.17139	&	823.86202	&	0.02386	\\
55540.609774	&	23.02603	&	235.38125	&	2412.73296	&	2648.11421	&	0.04814	\\
55540.952368	&	5.51462	&	117.68285	&	765.02534	&	882.70819	&	0.01524	\\
55540.963266	&	18.93492	&	117.69062	&	2236.19702	&	2353.88765	&	0.02537	\\
55541.343321	&	3.94267	&	117.70013	&	765.00806	&	882.70819	&	0.01674	\\
55541.395084	&	14.23128	&	411.93533	&	2648.12371	&	3060.05904	&	0.01220	\\
55542.043493	&	1.26695	&	58.85395	&	58.84618	&	117.70013	&	0.02280	\\
55542.082316	&	7.81970	&	176.53594	&	1294.64438	&	1471.18032	&	0.02129	\\
55542.610851	&	26.84851	&	529.62682	&	2883.52397	&	3413.15078	&	0.03662	\\
55542.691222	&	7.77636	&	117.69926	&	882.70819	&	1000.40746	&	0.02239	\\
55542.725277	&	4.57177	&	58.84531	&	176.53680	&	235.38211	&	0.04254	\\
55543.163907	&	14.31275	&	411.92669	&	1059.25363	&	1471.18032	&	0.03769	\\
55544.722950	&	7.26750	&	176.52816	&	588.48077	&	765.00893	&	0.03342	\\
55545.570242	&	5.29618	&	176.54544	&	823.86374	&	1000.40918	&	0.01745	\\
55545.611108	&	16.07877	&	235.39075	&	1176.94598	&	1412.33674	&	0.05000	\\
55547.527733	&	35.88220	&	588.46435	&	2530.43914	&	3118.90349	&	0.02779	\\
55548.411805	&	6.62462	&	176.53680	&	706.17398	&	882.71078	&	0.01821	\\
55548.777558	&	42.66308	&	117.69149	&	2000.81146	&	2118.50294	&	0.09381	\\
55549.246839	&	14.48484	&	117.70013	&	2295.04061	&	2412.74074	&	0.01299	\\
55550.629480	&	18.35971	&	470.78237	&	2236.19616	&	2706.97853	&	0.01681	\\
55552.452797	&	10.20249	&	353.08310	&	882.72029	&	1235.80339	&	0.03406	\\
56419.872987	&	42.42996	&	235.39507	&	3295.63123	&	3531.02630	&	0.07203	\\
56421.178735	&	18.79582	&	176.54630	&	1647.81994	&	1824.36624	&	0.02747	\\
56422.598915	&	59.26194	&	2707.13491	&	2883.67258	&	5590.80749	&	0.02899	\\
\hline
\end{tabular}
\end{center}
\end{table*}

\setcounter{table}{3}
\begin{table*}
\begin{center}
\caption{Parameters derived from the OPEA model by least squares method.}
\begin{tabular}{lcr}
\hline
\hline
Parameter	&	Values	&	95$\%$ Confidence Intervals	\\
\hline	 	 	 	 	 
$y_{0}$	&	0.34741	&	0.2440 to 0.4508	\\
$Plateau$	&	2.3121	&	2.1232 to 2.5010	\\
$K$	&	0.00031033	&	0.00024245 to 0.00037821	\\
$Tau$	&	3222.4	&	2644.0 to 4124.6	\\
$Half-time$	&	2233.6	&	1832.7 to 2859.0	\\
$Span$	&	1.9647	&	1.7914 to 2.1380	\\
$R^{2}$	&	- 	&	0.75	\\
\hline
\end{tabular}
\end{center}
\end{table*}

\setcounter{table}{4}
\begin{table*}
\begin{center}
\caption{Minima times and their residuals.}
\begin{tabular}{cccccc}
\hline
\hline
MJD (Obs)	&	$E$	&	Type	&	$O-C (I)$ (day)	&	$O-C (II)$ (day)	&	linear fit to the $O-C (I)$	\\
\hline
55373.637352	&	296	&	I	&	0.00175	&	0.00045	&	0.00130	\\
55375.015258	&	297	&	I	&	0.00100	&	-0.00030	&	0.00130	\\
55376.394783	&	298	&	I	&	0.00188	&	0.00058	&	0.00130	\\
55377.772232	&	299	&	I	&	0.00068	&	-0.00063	&	0.00130	\\
55379.151475	&	300	&	I	&	0.00127	&	-0.00004	&	0.00130	\\
55380.530204	&	301	&	I	&	0.00135	&	0.00004	&	0.00131	\\
55381.909019	&	302	&	I	&	0.00151	&	0.00020	&	0.00131	\\
55383.287462	&	303	&	I	&	0.00130	&	0.00000	&	0.00131	\\
55384.665830	&	304	&	I	&	0.00102	&	-0.00029	&	0.00131	\\
55386.045957	&	305	&	I	&	0.00250	&	0.00119	&	0.00131	\\
55387.423309	&	306	&	I	&	0.00120	&	-0.00011	&	0.00131	\\
55388.801861	&	307	&	I	&	0.00110	&	-0.00021	&	0.00132	\\
55390.180721	&	308	&	I	&	0.00131	&	-0.00001	&	0.00132	\\
55392.937562	&	310	&	I	&	0.00085	&	-0.00047	&	0.00132	\\
55394.316284	&	311	&	I	&	0.00092	&	-0.00040	&	0.00132	\\
55397.074523	&	313	&	I	&	0.00186	&	0.00054	&	0.00133	\\
55398.452606	&	314	&	I	&	0.00130	&	-0.00003	&	0.00133	\\
55401.209976	&	316	&	I	&	0.00136	&	0.00003	&	0.00133	\\
55402.588586	&	317	&	I	&	0.00133	&	-0.00001	&	0.00133	\\
55403.967495	&	318	&	I	&	0.00158	&	0.00025	&	0.00133	\\
55405.345561	&	319	&	I	&	0.00100	&	-0.00034	&	0.00134	\\
55406.723591	&	320	&	I	&	0.00038	&	-0.00096	&	0.00134	\\
55408.102754	&	321	&	I	&	0.00089	&	-0.00045	&	0.00134	\\
55409.481348	&	322	&	I	&	0.00084	&	-0.00051	&	0.00134	\\
55410.860243	&	323	&	I	&	0.00108	&	-0.00026	&	0.00134	\\
55412.238990	&	324	&	I	&	0.00118	&	-0.00017	&	0.00135	\\
55413.617981	&	325	&	I	&	0.00152	&	0.00017	&	0.00135	\\
55414.997250	&	326	&	I	&	0.00214	&	0.00079	&	0.00135	\\
55416.374600	&	327	&	I	&	0.00084	&	-0.00051	&	0.00135	\\
55417.753642	&	328	&	I	&	0.00123	&	-0.00012	&	0.00135	\\
55419.132621	&	329	&	I	&	0.00156	&	0.00020	&	0.00135	\\
55420.510400	&	330	&	I	&	0.00069	&	-0.00067	&	0.00136	\\
55421.888921	&	331	&	I	&	0.00056	&	-0.00080	&	0.00136	\\
55426.025010	&	334	&	I	&	0.00069	&	-0.00067	&	0.00136	\\
55427.403240	&	335	&	I	&	0.00027	&	-0.00109	&	0.00136	\\
55428.783068	&	336	&	I	&	0.00145	&	0.00009	&	0.00137	\\
55430.160693	&	337	&	I	&	0.00043	&	-0.00094	&	0.00137	\\
55432.918294	&	339	&	I	&	0.00073	&	-0.00064	&	0.00137	\\
\hline
\end{tabular}
\end{center}
\end{table*}

\setcounter{table}{4}
\begin{table*}
\begin{center}
\caption{Continued.}
\begin{tabular}{cccccc}
\hline
\hline
MJD (Obs)	&	$E$	&	Type	&	$O-C (I)$ (day)	&	$O-C (II)$ (day)	&	linear fit to the $O-C (I)$	\\
\hline
55434.297248	&	340	&	I	&	0.00103	&	-0.00034	&	0.00137	\\
55435.675787	&	341	&	I	&	0.00092	&	-0.00045	&	0.00137	\\
55437.054380	&	342	&	I	&	0.00086	&	-0.00051	&	0.00138	\\
55438.433266	&	343	&	I	&	0.00110	&	-0.00028	&	0.00138	\\
55439.811817	&	344	&	I	&	0.00100	&	-0.00038	&	0.00138	\\
55441.189359	&	345	&	I	&	-0.00011	&	-0.00149	&	0.00138	\\
55442.567810	&	346	&	I	&	-0.00031	&	-0.00169	&	0.00138	\\
55443.947869	&	347	&	I	&	0.00110	&	-0.00028	&	0.00139	\\
55445.326462	&	348	&	I	&	0.00104	&	-0.00034	&	0.00139	\\
55446.704701	&	349	&	I	&	0.00063	&	-0.00076	&	0.00139	\\
55448.082487	&	350	&	I	&	-0.00023	&	-0.00162	&	0.00139	\\
55449.463059	&	351	&	I	&	0.00169	&	0.00030	&	0.00139	\\
55450.841227	&	352	&	I	&	0.00121	&	-0.00019	&	0.00139	\\
55452.221730	&	353	&	I	&	0.00306	&	0.00166	&	0.00140	\\
55453.598523	&	354	&	I	&	0.00120	&	-0.00019	&	0.00140	\\
55454.977153	&	355	&	I	&	0.00118	&	-0.00022	&	0.00140	\\
55456.356654	&	356	&	I	&	0.00203	&	0.00063	&	0.00140	\\
55457.734853	&	357	&	I	&	0.00158	&	0.00018	&	0.00140	\\
55459.113623	&	358	&	I	&	0.00170	&	0.00030	&	0.00140	\\
55460.492162	&	359	&	I	&	0.00159	&	0.00018	&	0.00141	\\
55461.872023	&	360	&	I	&	0.00280	&	0.00139	&	0.00141	\\
55463.249164	&	361	&	I	&	0.00129	&	-0.00012	&	0.00141	\\
55464.627459	&	362	&	I	&	0.00094	&	-0.00047	&	0.00141	\\
55466.006470	&	363	&	I	&	0.00130	&	-0.00012	&	0.00141	\\
55467.385217	&	364	&	I	&	0.00139	&	-0.00002	&	0.00141	\\
55468.763258	&	365	&	I	&	0.00078	&	-0.00063	&	0.00142	\\
55470.141569	&	366	&	I	&	0.00045	&	-0.00097	&	0.00142	\\
55471.521176	&	367	&	I	&	0.00140	&	-0.00002	&	0.00142	\\
55472.899686	&	368	&	I	&	0.00126	&	-0.00016	&	0.00142	\\
55474.278113	&	369	&	I	&	0.00104	&	-0.00038	&	0.00142	\\
55475.656836	&	370	&	I	&	0.00111	&	-0.00031	&	0.00142	\\
55477.036061	&	371	&	I	&	0.00169	&	0.00026	&	0.00143	\\
55478.414494	&	372	&	I	&	0.00147	&	0.00004	&	0.00143	\\
55479.792703	&	373	&	I	&	0.00103	&	-0.00040	&	0.00143	\\
55481.170834	&	374	&	I	&	0.00051	&	-0.00092	&	0.00143	\\
55482.549788	&	375	&	I	&	0.00081	&	-0.00062	&	0.00143	\\
55483.929219	&	376	&	I	&	0.00159	&	0.00016	&	0.00144	\\
55485.307269	&	377	&	I	&	0.00099	&	-0.00044	&	0.00144	\\
55486.685532	&	378	&	I	&	0.00061	&	-0.00083	&	0.00144	\\
55488.063580	&	379	&	I	&	0.00000	&	-0.00144	&	0.00144	\\
\hline
\end{tabular}
\end{center}
\end{table*}

\setcounter{table}{4}
\begin{table*}
\begin{center}
\caption{Continued.}
\begin{tabular}{cccccc}
\hline
\hline
MJD (Obs)	&	$E$	&	Type	&	$O-C (I)$ (day)	&	$O-C (II)$ (day)	&	linear fit to the $O-C (I)$	\\
\hline
55489.442769	&	380	&	I	&	0.00054	&	-0.00090	&	0.00144	\\
55490.824094	&	381	&	I	&	0.00322	&	0.00177	&	0.00144	\\
55492.200729	&	382	&	I	&	0.00120	&	-0.00024	&	0.00145	\\
55494.959266	&	384	&	I	&	0.00244	&	0.00099	&	0.00145	\\
55496.337138	&	385	&	I	&	0.00166	&	0.00021	&	0.00145	\\
55496.337138	&	385	&	I	&	0.00166	&	0.00021	&	0.00145	\\
55497.716000	&	386	&	I	&	0.00187	&	0.00042	&	0.00145	\\
55499.095410	&	387	&	I	&	0.00263	&	0.00118	&	0.00145	\\
55499.095410	&	387	&	I	&	0.00263	&	0.00118	&	0.00145	\\
55500.473644	&	388	&	I	&	0.00222	&	0.00076	&	0.00146	\\
55501.851611	&	389	&	I	&	0.00153	&	0.00007	&	0.00146	\\
55503.230158	&	390	&	I	&	0.00143	&	-0.00003	&	0.00146	\\
55504.609336	&	391	&	I	&	0.00196	&	0.00049	&	0.00146	\\
55505.987369	&	392	&	I	&	0.00134	&	-0.00012	&	0.00146	\\
55507.367066	&	393	&	I	&	0.00239	&	0.00092	&	0.00146	\\
55508.745117	&	394	&	I	&	0.00179	&	0.00032	&	0.00147	\\
55511.503144	&	396	&	I	&	0.00251	&	0.00104	&	0.00147	\\
55512.881040	&	397	&	I	&	0.00176	&	0.00029	&	0.00147	\\
55514.259500	&	398	&	I	&	0.00157	&	0.00010	&	0.00147	\\
55515.637647	&	399	&	I	&	0.00107	&	-0.00041	&	0.00148	\\
55517.016941	&	400	&	I	&	0.00171	&	0.00023	&	0.00148	\\
55518.395944	&	401	&	I	&	0.00206	&	0.00058	&	0.00148	\\
55519.774024	&	402	&	I	&	0.00149	&	0.00001	&	0.00148	\\
55521.153793	&	403	&	I	&	0.00261	&	0.00113	&	0.00148	\\
55522.532080	&	404	&	I	&	0.00225	&	0.00076	&	0.00148	\\
55525.288367	&	406	&	I	&	0.00123	&	-0.00025	&	0.00149	\\
55526.668079	&	407	&	I	&	0.00230	&	0.00081	&	0.00149	\\
55528.046162	&	408	&	I	&	0.00173	&	0.00024	&	0.00149	\\
55529.425603	&	409	&	I	&	0.00252	&	0.00103	&	0.00149	\\
55530.802659	&	410	&	I	&	0.00092	&	-0.00057	&	0.00149	\\
55533.561410	&	412	&	I	&	0.00238	&	0.00088	&	0.00150	\\
55534.939002	&	413	&	I	&	0.00132	&	-0.00018	&	0.00150	\\
55536.319211	&	414	&	I	&	0.00288	&	0.00137	&	0.00150	\\
55537.697510	&	415	&	I	&	0.00252	&	0.00102	&	0.00150	\\
55539.076038	&	416	&	I	&	0.00240	&	0.00090	&	0.00150	\\
55540.453980	&	417	&	I	&	0.00169	&	0.00019	&	0.00151	\\
55541.833100	&	418	&	I	&	0.00216	&	0.00066	&	0.00151	\\
55543.212646	&	419	&	I	&	0.00306	&	0.00155	&	0.00151	\\
55544.589008	&	420	&	I	&	0.00077	&	-0.00074	&	0.00151	\\
55545.968579	&	421	&	I	&	0.00169	&	0.00018	&	0.00151	\\
\hline
\end{tabular}
\end{center}
\end{table*}

\setcounter{table}{4}
\begin{table*}
\begin{center}
\caption{Continued.}
\begin{tabular}{cccccc}
\hline
\hline
MJD (Obs)	&	$E$	&	Type	&	$O-C (I)$ (day)	&	$O-C (II)$ (day)	&	linear fit to the $O-C (I)$	\\
\hline
55547.346980	&	422	&	I	&	0.00144	&	-0.00007	&	0.00151	\\
55548.726156	&	423	&	I	&	0.00197	&	0.00045	&	0.00152	\\
55550.104937	&	424	&	I	&	0.00210	&	0.00058	&	0.00152	\\
55551.483063	&	425	&	I	&	0.00158	&	0.00006	&	0.00152	\\
56420.033207	&	1055	&	I	&	0.00206	&	-0.00055	&	0.00261	\\
56421.411327	&	1056	&	I	&	0.00153	&	-0.00108	&	0.00261	\\
56422.791294	&	1057	&	I	&	0.00285	&	0.00023	&	0.00261	\\
\hline
\end{tabular}
\end{center}
\end{table*}


\end{document}